\documentclass[12pt,a4paper]{article}
\usepackage[english]{babel}
\usepackage[square,sort,comma,numbers]{natbib}
\usepackage[utf8]{inputenc}
\usepackage[T1]{fontenc}
\usepackage{textcomp}
\usepackage{lmodern}
\usepackage[sc]{mathpazo}
\usepackage{hyperref}
\usepackage{a4wide}
\usepackage{amsmath}
\usepackage{gensymb}
\usepackage{graphicx}
\usepackage[labelfont=bf]{caption}
\usepackage{subcaption}
\usepackage[version=4]{mhchem}
\usepackage{makeidx}
\usepackage{fixltx2e}
\usepackage{float}
\usepackage{listings}
\usepackage{array}
\usepackage{tabu}
\usepackage{lineno}
\usepackage{setspace}
\DeclareUnicodeCharacter{2212}{-}
\def\SPSB#1#2{\rlap{\textsuperscript{{#1}}}\SB{#2}}
\def\SP#1{\textsuperscript{{#1}}}
\def\SB#1{\textsubscript{{#1}}}

\title{Atomistic Mechanism of the Nucleation of Methylammonium Lead Iodide Perovskite from Solution}
\author{Paramvir Ahlawat, Pablo Piaggi, Michael Graetzel, \\ Michele Parrinello and Ursula Rothlisberger{*}}
\date{}
\begin{document}
%\linenumbers
\maketitle
%\tableofcontents{}
%\chapter*{Abstract}
\section*{Abstract}
In the ongoing intense quest to increase the photoconversion efficiencies of lead halide perovskites, it has become evident that optimizing the morphology of the material is essential to achieve high peformance. Despite the fact that nucleation plays a key role in controlling the crystal morphology, very little is known about the nucleation and crystal growth processes. Here, we perform metadynamics simulations of nucleation of methylammonium lead triiodide (MAPI) in order to unravel the atomistic details of perovskite crystallization from a $\gamma$-butyrolactone solution. The metadynamics trajectories show that the nucleation process takes place in several stages. Initially, dense amorphous clusters mainly consisting of lead and iodide appear from the homogeneous solution. These clusters evolve into lead iodide (PbI\SB{2}) like structures. Subsequently, methylammonium (MA\SP{+}) ions diffuse into this PbI\SB{2}-like aggregates triggering the transformation into a perovskite crystal through a solid-solid transformation. Demonstrating the crucial role of the monovalent cations in crystallization, our simulations provide key insights into the evolution of the perovskite microstructure which is essential to make high-quality perovskite based solar cells and optoelectronics.

\newpage
%\section*{Main}
\section*{Introduction}
Hybrid organic-inorganic perovskites with the generic formula ABX\SB{3} (where A is a monovalent cation, i.e. methylammonium(MA\SP{+}), formamidinium(FA\SP{+}) or Cs\SP{+}, B is a divalent cation such as Pb\SP{2+} or Sn\SP{2+}, and X stands for I\SP{-}, Br\SP{-} or Cl\SP{-}) have shown remarkable optoelectronic properties with applications ranging from highly efficient low-priced solar cells \cite{Gratzel2014, Yang2017, Saliba2016a} to light-emitting diodes \cite{Tan2014} and photodetectors \cite{Fang2015}. Over the last years, it has been established that the performance of perovskite solar cells \cite{Sharenko2016, Salim2015, Bi2016, Yang2017}\cite{Snaith2018}, perovskites based LEDs and optoelectronics\cite{Huang2016, Akkerman2018} depends crucially on the morphology of the material\cite{Ball2016, Snaith2014, Eperon2014, Shi2015}. Various processing techniques\cite{Burschka2013, Im2011, KangningLiang1998} have been developed to obtain optimal results. These involve the use of additives \cite{Dar2014, Bi2016}, anti-solvents\cite{Jeon2014}, variations in the reaction temperature\cite{Dualeh2014} or the use of visible light \cite{Ummadisingu2017}\cite{Tsai2018} during nucleation. However, due to the lack of understanding of the crystallization process, these attempts have been confined to purely empirical trial-and-error strategies.

Some experimental studies have been able to elucidate few features of the crystallization mechanism. Several reports \cite{Moore2015, Manser2015, Rahimnejad2016, Stewart2016} have highlighted the importance of the lead halide coordination chemistry. In particular, based on the Tyndall effect, Yan \textit{et al.} \cite{Yan2015} found that nucleation takes a non-trivial route through the formation of colloids of various Pb-X complexes namely [PbX\SB{3}]\SP{-}, [PbX\SB{4}]\SP{2-}, [PbX\SB{5}]\SP{3-} and [PbX\SB{6}]\SP{4-}. Furthermore, the x-ray diffraction (XRD) and dynamic light scattering analysis by McMeekin \textit{et al.} \cite{McMeekin2017} suggested that such colloidal clusters serve as nucleation sites. Recently, Hu and co-workers \cite{Hu2017} investigated the crystallization mechanism of perovskites with in-situ synchrotron-based grazing incidence XRD. They found an intermediate phase containing [PbI\SB{6}]\SP{4-} that precedes the formation of the perovskite crystal. Moreover, several experimental studies\cite{Ummadisingue1, Bi2016} have demonstrated that the structural evolution of the intermediate adducts plays a key role in determining the ultimate performance of perovskites\cite{Stewart2016}.

All of these studies suggest that crystallization proceeds via the formation of different complex intermediates. However, the nature and formation of these intermediate phases and their conversion into the perovskite structure have not been established. Knowledge about these fundamental aspects of crystallization would likely allow for a more rational control over the perovskite microstructure and thereby to materials with improved photoconversion efficiencies.

Molecular dynamics (MD) simulations can in principle provide in-depth atomistic level details of nucleation processes\cite{Salvalaglio2015}\cite{Salvalaglio2015a}. However, simulating the nucleation of organic-inorganic lead halide perovskites from solution is a very challenging problem. In fact, MD simulations on nucleation of such a complex hybrid organic-inorganic system have never been performed before. For one thing, due to the presence of several species, competing crystalline phases may exist and the nucleation pathway can be highly non-trivial. For another, nucleation is a rare event and the time scales of this process typically exceed the affordable times in standard MD simulations. To overcome the time scale limitation, in this work we employ well-tempered metadynamics (WTMetaD)\cite{Laio2002}\cite{Barducci2008}. 

Methylammonium lead triiodide (MAPI) has been the champion of halide perovskite materials and is the current most studied among this class of materials. We perform WTMetaD simulations to study the nucleation mechanism of MAPI from a $\gamma$-butyrolactone (GBL) solution which is one of the most commonly used solvents for the crystallization of lead halide perovskites \cite{Saidaminov2015, Jeon2014, Heo2013, Im2011, Kadro2015,Zhang2017a}. Our WTMetaD simulations demonstrate that nucleation of MAPI from solution is a multi-step process. In this work, we describe the mechanism of the individual stages and present the atomistic details about the formation of intermediate phases and their conversion to the perovskite crystal. 

\section*{Results}
\begin{figure}
  \includegraphics[width=\linewidth]{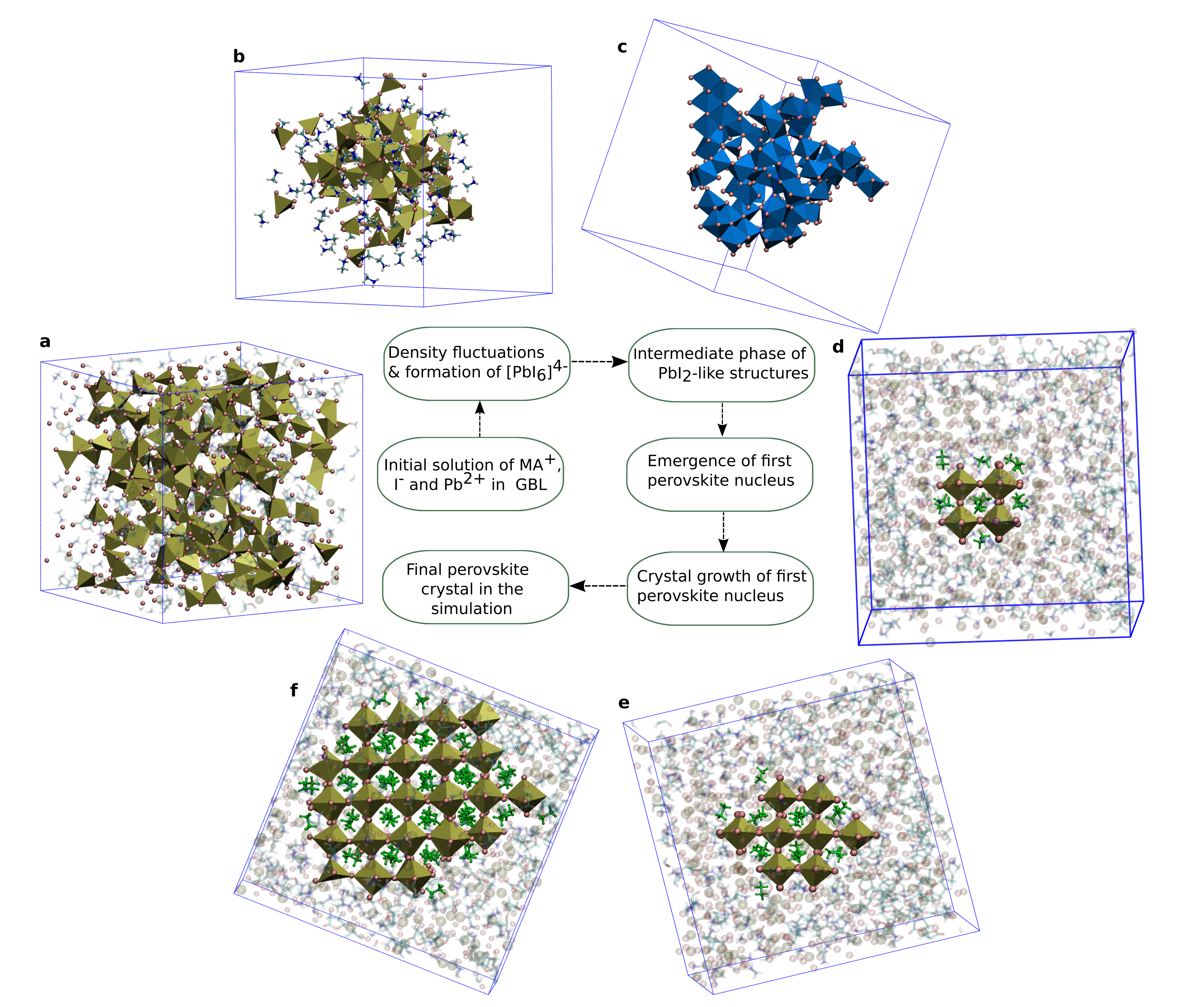}
  \caption{\textbf{Illustration of the full nucleation pathway.} Pb-I complexes are shown as golden and blue polyhedra with Pb\SP{2+} in the center and I\SP{-} on the corners. Free I\SP{-} is shown as pink spheres. Crystalline MA\SP{+} ions are shown in green. Snapshot (a) represents the initial solution of MA\SP{+}, I\SP{-} and Pb\SP{2+} in GBL. MA\SP{+} and GBL are shown semi-transparent to visualize the random distribution of Pb\SP{2+} and I\SP{-} in solution. Snapshot (b) represents the initial cluster formation of Pb\SP{2+} and I\SP{-} surrounded by MA\SP{+} ions. Snapshot (c) shows only the edge-sharing [PbI\SB{6}]\SP{4-} octahedra. Snapshot (d) shows the first perovskite nucleus observed in the solution. Snapshot (e) shows the growth of the initial nucleus. Snapshot (f) displays the largest perovskite crystal in the simulations. All of the images of these snapshots were generated with VMD-1.9.2 \cite{Humphrey1996}}
  \label{fig:full}
\end{figure}

We start from a well equilibrated solution where the MA\SP{+}, Pb\SP{2+} and I\SP{-} ions are initially homogeneously distributed in GBL. We have prepared two diferent simulation setups and will refer to them as simulation (A) and simulation (B). Simulation (B) has a lower concentration of solute species as compared to simulation (A). Further details are provided in the Methods section. In both cases different Pb-I complexes, such as [PbI\SB{2}], [PbI\SB{3}]\SP{-} and [PbI\SB{4}]\SP{2-} are formed spontaneously in these solutions with a uniform distribution over the solution. These inital configurations are shown in Fig. \ref{fig:full}a and the Supplementry Fig. S8a for simulation (A) and simulation (B), respectively.

As expected, during the course of standard (unbiased) MD simulations, no formation of a perovskite phase is observed due to the presence of a high free energy barrier. This free energy barrier can be rationalized on the basis of classical nucleation theory, which portrays the formation of crystalline clusters as an interplay between volume and surface contributions. In a macroscopic system, this leads to a free energy barrier for the formation of a cluster of a critical size. Before the critical cluster forms, the system undergoes order and density fluctuations that can be descibed as clusters of relatively small size. Eventually, a critical cluster is formed and the system is able to surmount the barrier. Further increase in the cluster size is spontaneous. At variance with this picture, in a confined system as the one used in our simulations, the solution becomes depleted of solute atoms as the cluster grows thus decreasing the supersaturation with time. As a consequence, a finite size cluster is not stable. However, this unwanted effect does not distort the crystallization path as previous experience has shown \cite{Salvalaglio2015}\cite{Salvalaglio2015a}. 

%To bring the formation of a perovskite crystal within affordable simulation time, 
To accelerate the rare nucleation process in such a way that it can be observed within our limited simulation time of the order of microseconds, we employ well-tempered metadynamics (WTMetaD) in order to enhance the thermal fluctuations. WTMetaD constructs an external bias potential as a function of a few collective coordinates also known as collective variables (CVs). This bias potential discourages frequently visited configurations thus enhancing the fluctuations of the CVs and helping the system to overcome free energy barriers. 

An appropriate choice of CVs is essential to the success of WTMetaD simulations. We have chosen two CVs to describe the nucleation process. The first (S\SB{P}) is the number of MA\SP{+} ions that have a coordination with MA\SP{+}, Pb\SP{2+} and I\SP{-} which is compatible with the perovskite phase. We also use S\SB{P} to represent the degree of crystallinity in the system. The second one (S\SB{PbI\SB{6}}) is the number of Pb\SP{2+} ions coordinated by six I\SP{-}. Further details about the CVs are described in the Methods section. 

\begin{figure}[H]
  \includegraphics[width=100mm,scale=1.0]{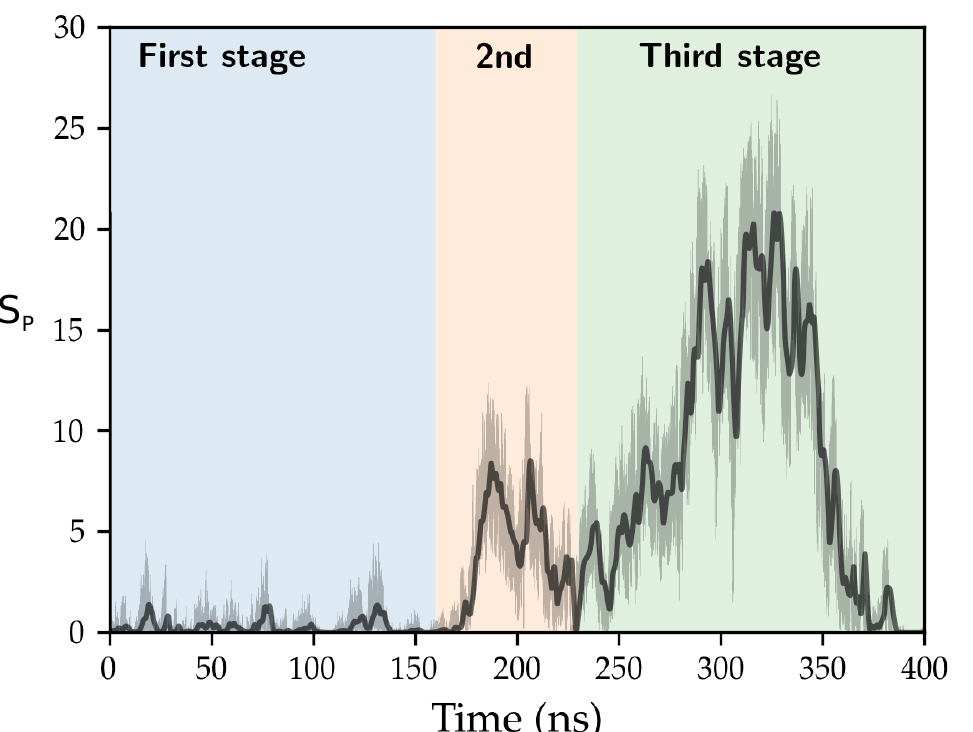}
  \caption{\textbf{Perovskite CV employed to drive the nucleation process.} In order to guide the eyes, the running average (over 1ns interval) of the CV is shown as a thick black line. The stages into which the simulation is divided for the purpose of analysis are shown with different colors (see text for details).}
  \label{fig:peroCVN}
\end{figure}

During the course of the WTMetaD simulations, perovskite crystals ultimately form in both simulations. This is also shown in Supplementary movies 1 and 3 for simulations (A). The formation of these large crystals is shown in Fig \ref{fig:full}f for simulation (A), in the Supplementary Fig. S8d for simulation (B). We have identified three different stages of the simulation trajectories that correspond to different steps during the nucleation and crystallization process as depicted in Fig \ref{fig:peroCVN} for simulation (A). At the first stage, amorphous clusters of Pb-I complexes emerge from the solution. In the second stage, the amorphous clusters partially transform into the perovskite structure. In the last stage of our trajectories, the crystal grows by sequential addition and arrangement of ions around the initial nucleus. In the next paragraphs, we describe each stage in detail for simulation (A). We have observed a similar qualitative nucleation mechanism in simulation (B) and the results of the latter are presented in the Supplementary Information (SI).

\begin{figure}[H]
  \includegraphics[width=100mm,scale=0.9]{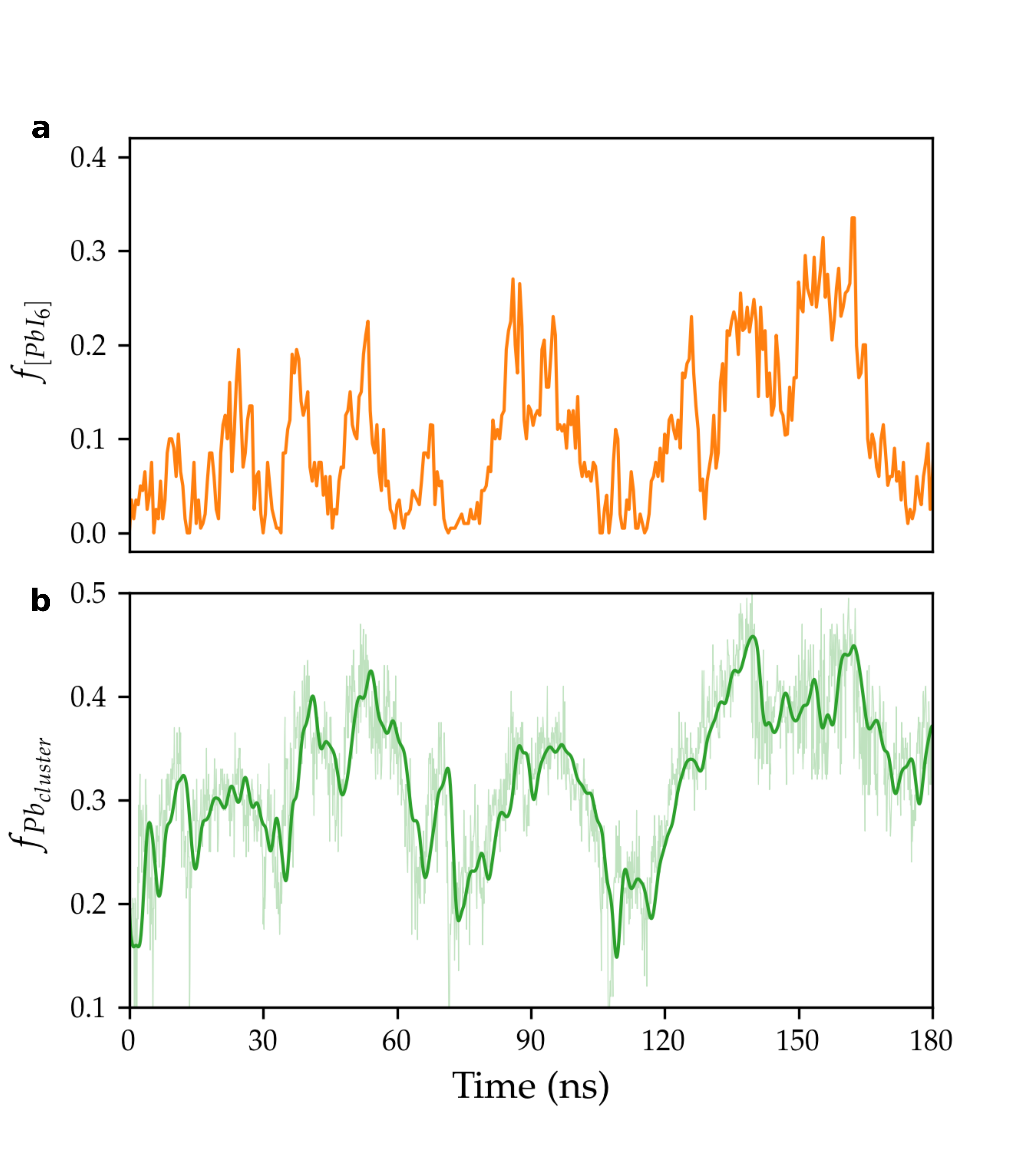}
  \caption{\textbf{Analysis of the first stage in which cluster formation is observed.} Top panel (a) display the the number of [PbI\SB{6}]\SP{4-} complexes divided by the total number of Pb\SP{2+} ions in our simulation. The bottom panel (b) shows the size of the largest cluster as a function of simulation time. Here we represent the size of the cluster as the fraction of Pb\SP{2+} ions in the cluster out of the total of Pb\SP{2+} ions in the simulation box.} 
  \label{fig:initial_cluster}
\end{figure}

At the first stage, we observe increased fluctuations in the number of [PbI\SB{6}]\SP{4-} octahedra as shown in Fig. \ref{fig:initial_cluster}a. The relative increase in the coordination of Pb\SP{2+} with I\SP{-} is achieved by two or more Pb\SP{2+} sharing the same I\SP{-}. As a result, clusters made from face or edge-sharing [PbI\SB{6}]\SP{4-} octahedra are formed. The time evolution of edge-sharing octahedra is presented in Supplementary Fig. S4. This is in contrast to the network of corner sharing [PbI\SB{6}]\SP{4-} octahedra that is characteristic for the final perovskite structure. MA\SP{+} ions surround the [PbI\SB{6}]\SP{4-} clusters to compensate the large negative charge. These processes increase the local density of solute species. From Fig. \ref{fig:initial_cluster}, we observe that eventually [PbI\SB{6}]\SP{4-} can transform back to lower order iodoplumbates but the high local density of solute species is preserved. In order to quantify the local density fluctuations, we have performed a clustering of Pb\SP{2+} ions that are coordinated by at least three I\SP{-} ions within a radial cutoff of 0.40 nm. Further details on the calculations of the size of these clusters are presented in the SI. At the end of this stage, after $\sim$ 140 ns, Pb\SP{2+} and I\SP{-} ions have formed relatively large amorphous clusters as depicted in Fig. \ref{fig:full}b. 

\begin{figure}
  \includegraphics[width=160mm,scale=1.0]{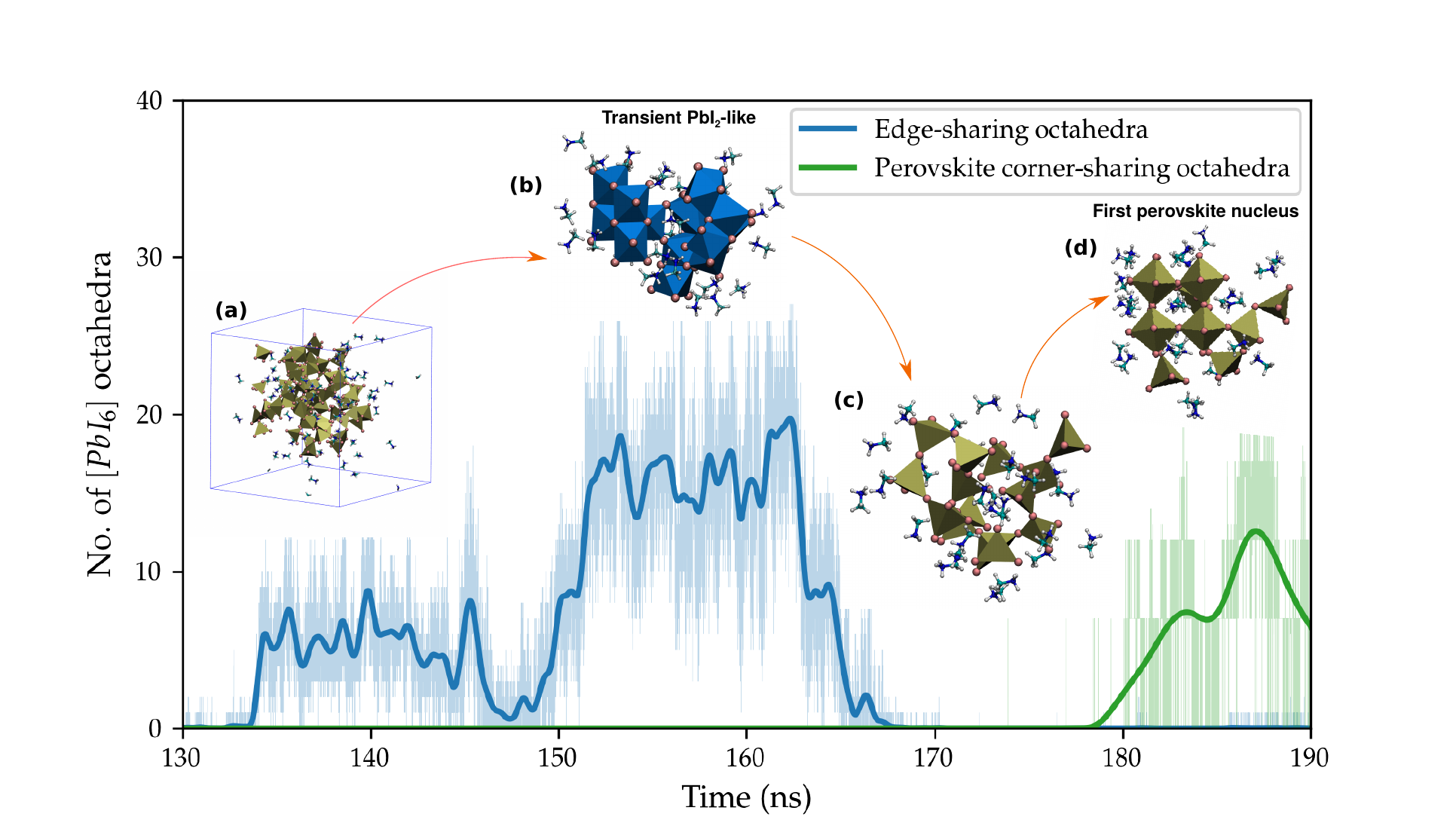}
  \caption{\textbf{Analysis of the second stage.} Time evolution of edge-sharing and perovskite corner-sharing octahedra. Thick lines are the running average (over 1ns interval) of the data and are shown in order to guide the eyes. Snapshot (a) shows the initial cluster at $\sim$ 140 ns. Snapshot (b) displays the configuration of the part of the cluster that evolves into the first perovskite nucleus at $\sim$ 161 ns. Snapshot (c) represents the evolution of the same cluster at $\sim$ 175 ns. Snapshot (d) displays the first perovskite nucleus in our simulation at $\sim$ 180 ns. All snapshots generated with VMD-1.9.2 \cite{Humphrey1996}}
  \label{fig:transition}
\end{figure}

\begin{figure}
  \includegraphics[width=80mm,scale=0.8]{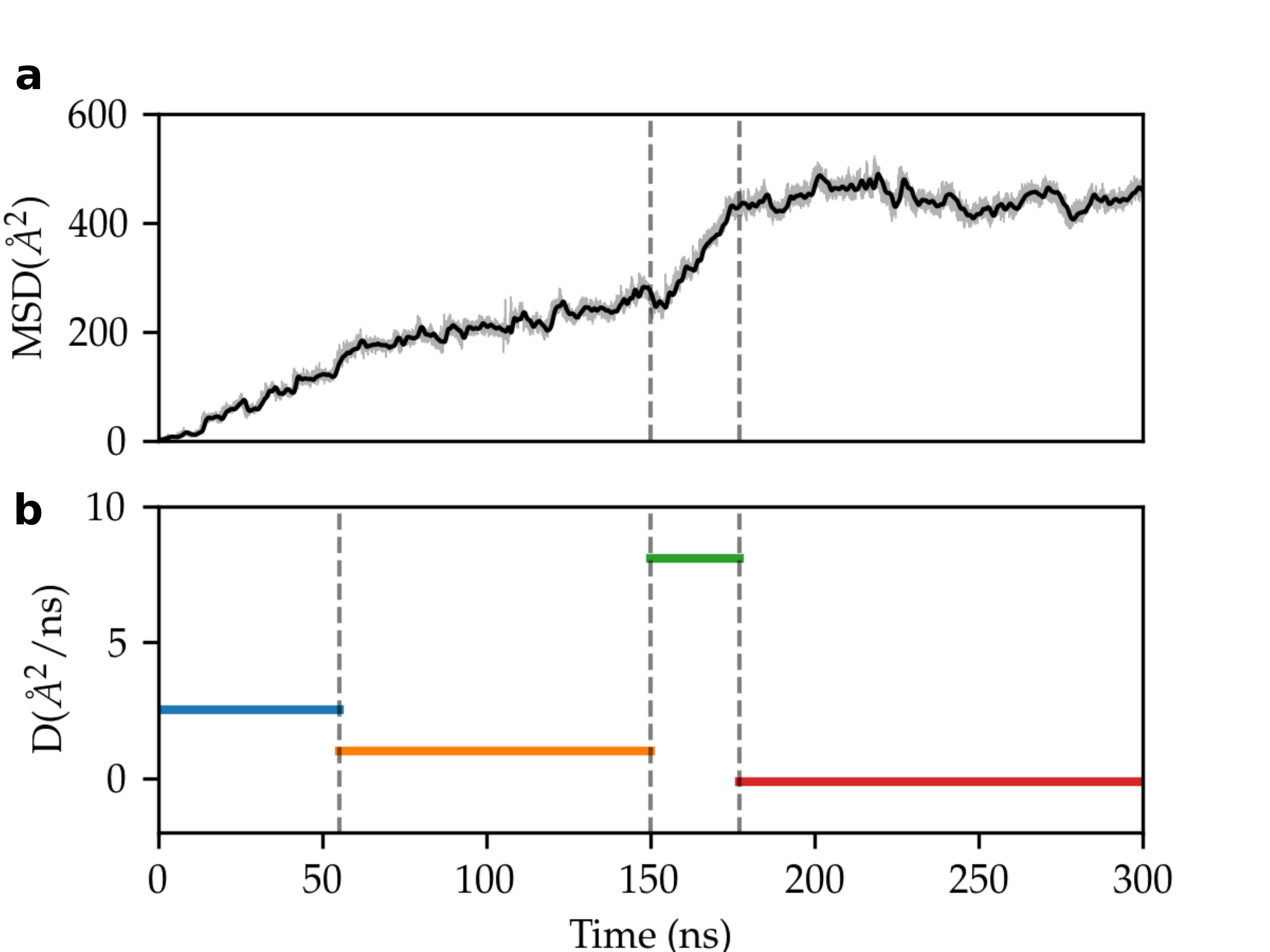}
  \caption{\textbf{Diffusion analysis of the subset of MA\SP{+} ions.} Time evolution of (a) mean square displacement and (b) diffusion coefficient of MA\SP{+} ions that take part in the formation of the first perovskite nucleus.}
  \label{fig:msd}
\end{figure}

In the next stage, a part of the amorphous clusters transforms into perovskite. It has been shown that PbI\SB{2}-like structures can be important intermediates (or byproducts) in the crystallization of MAPI \cite{Ummadisingue1, Jeon2014, Ahn2015}. Therefore, we quantify the formation of PbI\SB{2}-like structures in terms of the number of edge-sharing [PbI\SB{6}]\SP{4-} octahedra that share at least three edges with neighbouring [PbI\SB{6}]\SP{4-} octahedra. We also assess the amount of perovskite formed in the solution. To do so, we calculate the number of corner-sharing [PbI\SB{6}]\SP{4-} octahedra that share all six corners with neighbouring [PbI\SB{6}]\SP{4-} and Pb-I-Pb angles between them that are consistent with the cubic perovskite structure of MAPI at high temperature. Further details on selecting the angle cutoff and calculations of corner-sharing octahedra are provided in the SI. In Fig. \ref{fig:transition}, we show the number of edge-sharing and corner-sharing pervovskite octahedra as a function of simulation time. Initially, we observe that the number of edge-sharing octahedra increases and attains a maximum value at $\sim$ 161 ns. The configuration of edge-sharing octahedra at this time is shown in Fig. \ref{fig:transition}, Fig. \ref{fig:full}c and the Supplementary Fig. S3.  

From Supplementary Fig. S3 one can see that individual parts of these configurations show a remarkable resemblance to the rocksalt crystalline structure. However, these clusters are not big and are arranged in a way that leaves adequate space between them, therefore we qualitatively name the whole structure a PbI\SB{2}-like structure. Moreover, in the SI we show a detailed pictorial distribution of the clusters of PbI\SB{2}-like structures of edge-sharing octahedra. From the analysis presented in the Supplementary Fig. S1, it is evident that 95\% of the edge-sharing octahedra later convert to the perovskite structure. In the snapshots of Fig. \ref{fig:transition}, the time evolution of the ions that form the first perovskite structure is shown. 

Moving further along the trajectory the PbI\SB{2}-like structure starts to transform into [PbI\SB{4}]\SP{2-} tetrahedra. This can be seen in Fig \ref{fig:transition} where the number of edge-sharing octahedra decreases. Since the PbI\SB{2}-like structure opens up, space is created between Pb-I complexes. During this process, MA\SP{+} ions diffuse into these spaces between them. This mechanism can be observed in the snapshots of Fig. \ref{fig:transition}. To quantify the diffusion of MA\SP{+} ions, we have calculated the selfdiffusion coefficient (D) of the subset of MA\SP{+} ions that form the first perovskite crystal. From Fig. \ref{fig:msd}, a sharp increase in the diffusion of the MA\SP{+} ions from $\sim$ 155 to $\sim$ 177 ns can be noticed. The sharp increase in D of MA\SP{+} ions takes place during the time interval where PbI\SB{2}-like structures are the most abundant. This suggests that the formation of PbI\SB{2} seems to be correlated with the diffusion of MA\SP{+} ions inside the Pb-I clusters. From snapshot (c) in Fig. \ref{fig:transition}, one can see that MA\SP{+} ions are located in the vicinity of [PbI\SB{4}]\SP{2-} tetrahedra. 

Around 180 ns this structure rearranges into perovskite corner-sharing [PbI\SB{6}]\SP{4-} octahedra with MA\SP{+} ions in the center of cubo-octahedral cages. A snapshot of the first born perovskite crystal is shown in Fig.\ref{fig:transition}. It can also be seen in Fig. \ref{fig:transition} that there is an increase in the number of perovskite corner-sharing octahedra compatible with the formation of a perovskite phase. Also, as seen in Fig.\ref{fig:msd}, the diffusion diminishes substantially implying that the MA\SP{+} ions have become part of the newly formed perovskite crystal. This whole process is also shown in Supplementary movie 2. 

\begin{figure}
  \includegraphics[width=170mm,scale=0.9]{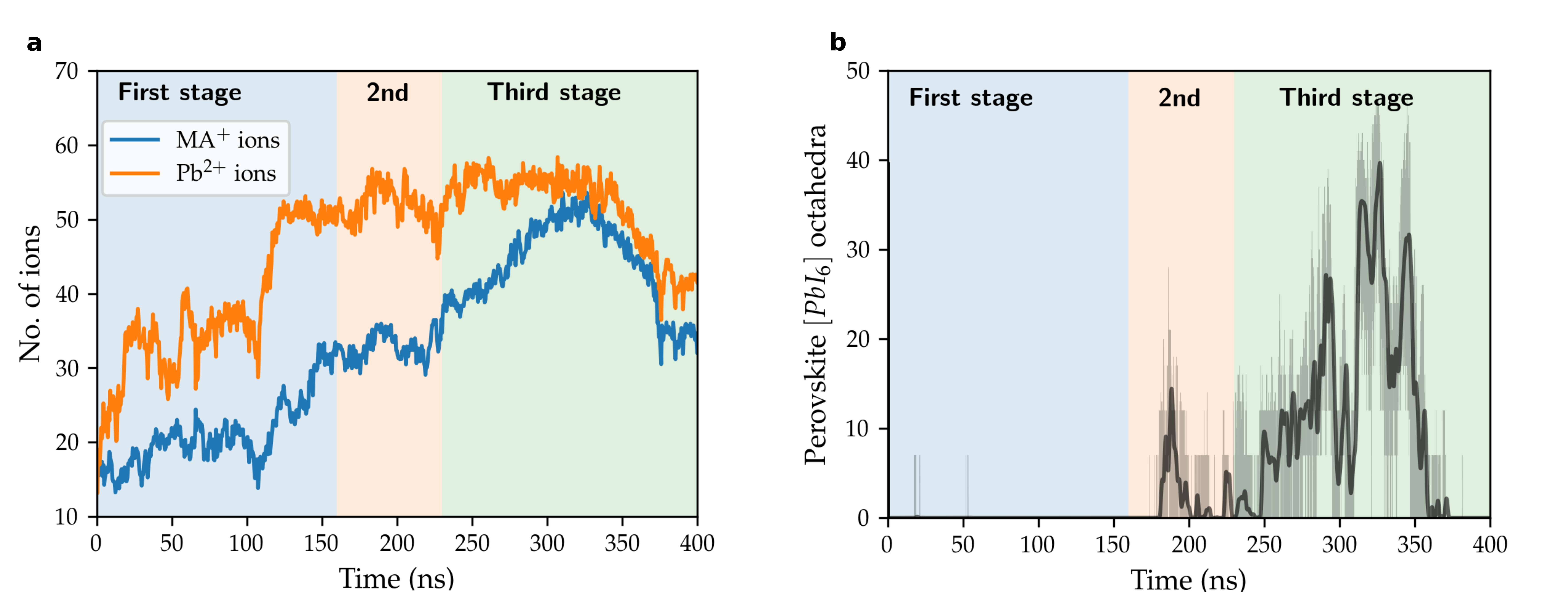}
  \label{fig:pero_oct}
  \caption{Left panel shows the change in concentration of MA\SP{+} and Pb\SP{2+} ions around the center of the first perovskite nucleus. Right panel displays the temporal evolution of perovskite corner-sharing octahedra. The thick black line in the right panel is the running average (over 1ns interval) of the data and is shown to guide the eyes.}
\end{figure}

At the final stage, the initial perovskite crystal grows as shown in Fig. \ref{fig:peroCVN}. Snapshots illustrating this process are displayed in Fig. \ref{fig:full}d-f. During this process, MA\SP{+} ions from solution come to the surface of the perovskite crystal. To quantify this, we select the MA\SP{+} and Pb\SP{2+} ions that form the largest perovskite crystal shown in Fig. \ref{fig:full}f and calculate the concentration of these ions in a specific region around the center of the first perovskite nucleus. The dimensions of this region are selected based on the size of the largest observed crystal corresponding to simulation time $\sim$ 327ns. We find that the concentration of Pb\SP{2+} ions at this stage remains approximately the same after the first stage. However, it can be observed from Fig. 6a that there is a continuous increase in the concentration of MA\SP{+} ions around the center of the nucleus and that reaches a maximum at $\sim$ 327ns. This strongly suggests that MA\SP{+} ions have a fundamental role in the nucleation process. As the MA\SP{+} ions reach the surface of the crystal, a simultaneous formation of perovskite corner-sharing octahedra around these MA\SP{+} ions occurs. This can be seen from the increase in the number of corner-sharing octahedra in Fig. 6b and the increase in $S\SB{P}$ in Fig. \ref{fig:peroCVN}. Around 327 ns, the crystal attains its maximum size and afterwards it gradually dissolves due to the above mentioned depletion effect in a finite size system.
%\ref{fig:MA_dens} \ref{fig:MA_dens} \ref{fig:pero_oct}
\section*{Discussion}

From these results, the following qualitative description of the nucleation process of MAPI perovskite from solution is emerging: Nucleation starts from the formation of aggregates of Pb\SP{2+} and I\SP{-}. This result is in general agreement with the experimental studies from Yan \textit{et al.}\cite{Yan2015} and Kadro \textit{et al.}\cite{Kadro2015} where colloids of Pb-I complexes display the Tyndall effect even at higher temperatures. 

Afterwards, these aggregates evolve into clusters of edge-sharing [PbI\SB{6}]\SP{4-} octahedra and a PbI\SB{2}-like intermediate phase is formed. Here we present the configuration of these intermediates and find that rocksalt type structure of edge-sharing octahedra dominate the PbI\SB{2}-like intermediate phase. Eventually, MA\SP{+} ions diffuse inside the Pb-I clusters and the PbI\SB{2}-like intermediate phase transforms into perovskite. In addition, we have identified that the continuous diffusion of MA\SP{+} ions into Pb-I clusters is key to the nucleation and growth of perovskite. Furthermore, our results suggest that the formation of a PbI\SB{2}-like phase might facilitate the diffusion of MA\SP{+} ions into the Pb-I clusters. 

These insights leads us to some generalized rules for controlling the nucleation process to optimize the shape and size of perovskite crystals. Perovskite thin films with larger grains are known to have high efficiency. From our simulations, we have found that controlling the diffusion of MA\SP{+} ions can in principle control the nucleation process. Therefore, by selecting typical solvents or additives that form strong intermediate interaction with MA\SP{+} ions, one can slow down the nucleation process to make larger grains or single crystals. Furthermore, lowering the concentration of MA\SP{+} ions in precursor solutions may result in larger nucleus size. Moreover, from our simulations, we can anticipate that the shape of perovskite crystals strongly depends on the structures of intermediate PbI\SB{2}-like phases. Although, our simulations does not clearly outline the different metastable structures of edge-sharing [PbI\SB{6}]\SP{4-} octahedra, however aggregates with rocksalt structure of these edge-sharing octahedra emerge as one of the main configuration. The evolution of intermediate structure formation might depend on solvents and additives, which can be further explored in future studies based on the methodology presented in this work.
%\newpage
%\newpage

\section*{Conclusions}
To conclude, in this work we employ WTMetaD simulations to uncover the salient atomistic features of the nucleation of MAPI from solution. This work constitutes a significant advancement in the field of MD simulations of nucleation of complex multicomponent systems from solution. These simulations reveal that a PbI\SB{2}-like phase appears from the clusters of Pb\SP{2+} and I\SP{-} before the onset of nucleation. This is followed by the diffusion of MA\SP{+} ions into PbI\SB{2}-like aggregates triggering the transformation into a perovskite. In this study, the relevant intermediate steps of the entire process were characterized with atomic detail. This mechanistic understanding of the nucleation process might help in the design of improved strategies to better control perovskite morphology to produce highly efficient perovskite based solar cells, LEDs, and photodetectors. Furthermore, our work lays the foundations to study the atomistic details of nucleation and crystal growth of different perovskites.
%From our simulations, we have discovered that the formation of [PbI\SB{6}]\SP{4-} brings the Pb-I species together and as a result amorphous clusters containing Pb\SP{2+} and I\SP{-} are formed.We have shown the complete details of structure reorganization of intermediate phases to form the perovskite structure
\newpage
\section*{Methods}
\subsection*{MD simulations}
Simulation (A) discussed in this work was prepared with 200 units of PbI\SB{2} and 200 units of MAI, which are randomly distributed in 112 GBL molecules by using the \textit{insert-molecule} utility of the gromacs simulation package. Simulation (B) was prepared in the same way with 180 units of PbI\SB{2} and 180 units MAI randomly distributed in 368 GBL molecules. A fixed point charge force field was chosen for MAPI \cite{Mattoni2015} that is known to reproduce the high temperature cubic phase and the generalized amber force field (GAFF)(http://virtualchemistry.org/) was used for GBL. The heterogeneous solute-solvent Van der Walls (VdW) parameters were calculated based on the mixing rules. In the force field for MAPI from Mattoni \textit{at el.}\cite{Mattoni2015}, the VdW parameters of Pb\SP{2+} and I\SP{-} are modelled with a Buckingham-type potential. Therefore to calculate the heterogeneous solute-solvent parameters with mixing rules, we used the lennard-jones parameters of Pb\SP{2+} and I\SP{-} from Gutierrez-Sevillano \textit{at el.}\cite{Gutierrez-Sevillano2015a}. We have chosen a 1.0 nm cutoff for nonbonded interactions and three-dimensional periodic boundary conditions were applied for each simulation. Long range electrostatic interactions are treated with the particle-particle-particle-mesh Ewald method. We employ the SHAKE algorithm\cite{Ryckaert1977} to constrain the bond length of hydrogen atoms. The time step used in all of the simulations is 2 fs. All simulations are performed with the Large-scale Atomic/Molecular Massively Parallel Simulator(LAMMPS) code (31 Mar 2017)\cite{Plimpton1995}. The systems were first minimized with a conjugate gradient algorithm with a tolerance of maximum residual force of 10\SP{-3} kcal/mole-\AA. After minimization, the systems were relaxed with a 100ps NVT equilibration run. All production run simulations presented in this work were carried out in the isothermal-isobaric ensemble. We use a velocity rescaling thermostat \cite{Bussi2007} with a relaxation time of 0.1 ps. The Parrinello-Rahman barostat \cite{Parrinello1981} was used to keep the pressure equal to the standard atmospheric pressure. The relaxation time of the barostat was set to 10ps. With this setup, at first we equilibrate the fully solvated simulations for 500 ps. In this way, the ions are uniformly dispersed in the solvent and density inhomogeneities are eliminated. After completing the equilibration run, the production WTmetaD run was started for longer time scales, typically up to 400 ns. WTmetaD simulations were performed with a set of CVs which are described in detail in the next section. 

%The heterogeneous solute-solvent Van der Walls (VdW) parameters were calculated based on the mixing-rule. In force field for MAPI from Mattoni \textit{at el.}\cite{Mattoni2015}, the VdW parameters of Pb\SP{2+} and I\SP{-} are modeled with Buckingham-type potential. Therefore to calculate the heterogeneous solute-solvent parameters with mixing rule, here we have used the lennard-jones parameters of Pb\SP{2+} and I\SP{-} from Gutierrez-Sevillano \textit{at el.}\cite{Gutierrez-Sevillano2015}.

\subsection*{Collective variables}
The choice of the CVs is critical to be able to describe the crystallization from enhanced sampling simulations such as WT-MetaD. It is essential to carefully select the appropriate collective variables (CVs) which can enhance the fluctuations in the local order of crystalline species that form in solution. A multitude of CVs have been used to study crystallization, however most of them can only force the system to crystallize in specific crystalline structures. In this work, we propose a new CV named \textit{S\SB{P}}, that does not presume a specific crystalline phase of MAPI. Thus, it enables us to observe all complexes that form before the onset of nucleation and reveal important details about the early stages of nucleation. This CV describes the local order of the species as a product of the individual local densities of MA\SP{+}, Pb\SP{2+} and I\SP{-} centers around MA\SP{+}. Here, we represent MA\SP{+} with the position of the center of C-N axis of the MA\SP{+} ion. At first, we represent the local densities in the form of coordination numbers ($f_i^{\alpha}$) of individual species MA\SP{+}, Pb\SP{+} and I\SP{-} surrounding the position \textit{i} in the center of C-N axis of MA\SP{+} within a defined cutoff radius ($r^{\alpha}_{cut}$). 
\begin{equation}\label{eqn:CV11}
f_i^{\alpha} = \sum_{j=1}^{N_{\alpha}} \frac{ 1 - \left(\frac{r_{ij}}{r^{\alpha}_{cut}}\right)^a } 
{ 1 - \left(\frac{r_{ij}}{r^{\alpha}_{cut}}\right)^b}
\end{equation}
Where $r_{ij}$ is the distance between positions of \textit{i} and \textit{j}. The values for $r^{\alpha}_{cut}$ are selected based on the first minium of the radial distribution function (g(r)) of MA\SP{+} in the perovskite structure. To calculate the local densities, $\rho_i^{\alpha}$, we define another switching function in the space of coordination numbers. 
\begin{equation}\label{eqn:CV12}
\rho_i^{\alpha} = \frac{ 1 - \left(\frac{f_i^{\alpha}}{f^{\alpha}_{cut}}\right)^{-m}} 
{1 - \left(\frac{f_i^{\alpha}}{f^{\alpha}_{cut}}\right)^{-n}}
\end{equation}
Where $f^{\alpha}_{cut}$ is the lower threshold of the coordination number. The values for $f^{\alpha}_{cut}$ are chosen equal to the number of MA\SP{+}, Pb\SP{2+} and I\SP{-} in the first coordination sphere of MA\SP{+} in the perovskite structure of MAPI. The local order around \textit{i} is calculated by multiplying the individual local densities of each species. 
\begin{equation}\label{eqn:CV12}
\eta_i=\prod_{\alpha = MA, Pb, I} \rho_i^{\alpha} 
\end{equation}
The final expression for the CV becomes the following:
\begin{equation}\label{eqn:CV14}
S_{P}= \sum_{i=1}^{N_{MA}} \eta_i
\end{equation}
Where $N_{MA}$ is the number of MA\SP{+} ions around which the variables are defined. We have implemented this CV in PLUMED-2.4\cite{Tribello2014}. The values for $r^{\alpha}_{cut}$ and $f^{\alpha}_{cut}$ that are used in our simulation are shown in the following table.
\newline
\newline
\begin{minipage}{\linewidth}
\centering
\captionof{table}{Values for the switching functions used in \textit{S\SB{P}}} \label{tab:CV_table} 
\begin{tabu} to 0.8\textwidth { | X[c] | X[c] | X[c] | }
 \hline
 Species & $r^{\alpha}_{cut}$ & $f^{\alpha}_{cut}$ \\
 \hline
 MA\SP{+} - MA\SP{+}  & 0.75nm  & 6.0 \\
 \hline
 MA\SP{+} - Pb\SP{2+} & 0.70nm & 8.0 \\
 \hline
 MA\SP{+} - I\SP{-}  & 0.55nm  & 12.0 \\
 \hline
\end{tabu}\par
\bigskip
\bigskip
\end{minipage}
\newline
\newline
The second CV, \textit{S\SB{PbI\SB{6}}} is defined as the number of Pb\SP{2+} ions that have a coordination of six with I\SP{-} within a certain threshold. At first, we calculate the coordination number of Pb\SP{2+} with I\SP{-}.
\begin{equation}\label{eqn:CV21}
f_i^{Pb-I} = \sum_{j=1}^{N_{I}} \frac{ 1 - \left(\frac{r_{ij}}{r^{Pb-I}_{cut}}\right)^a } 
{ 1 - \left(\frac{r_{ij}}{r^{Pb-I}_{cut}}\right)^b}
\end{equation}
Where $r_{ij}$ is the distance between positions of \textit{i} and \textit{j}. Here \textit{i} and \textit{j} represent the indices of Pb\SP{2+} and I\SP{-}. Then we define another switching function in the space of coordination number of Pb-I. 
\begin{equation}\label{eqn:CV22}
S_{PbI_6} = \sum_{i}^{N_{Pb}}\frac{ 1 - \left(\frac{f_i^{Pb-I}}{f^{Pb-I}_{cut}}\right)^{-m}} 
{1 - \left(\frac{f_i^{Pb-I}}{f^{Pb-I}_{cut}}\right)^{-n}}
\end{equation}
The values for \textit{r\SPSB{Pb-I}{cut}} and \textit{f{\SPSB{Pb-I}{cut}}} are taken as 0.39 nm and 6. The height of the Gaussian bias potentials is set to $\approx$ 2k\SB{B}T. The width of the Gaussian bias potentials is set to 0.5 and 1.0 for  \textit{S\SB{P}} and \textit{S\SB{PbI\SB{6}}} respectively. The Gaussian hills are deposited after every 1 ps. We have chosen a bias factor of 300 that allows for an extensive exploration of hundreds of solute molecules contained in a typical simulation box. More theoretical details on WTmetaD can be found in references\cite{Barducci2008}\cite{Laio2002}.

%for $S_{Pero}$ and $S_{PbI_6}$ 

\subsection*{Code availability}
All of the codes used in this study such as the calculations of the reaction coordinate $S_{P}$, edge-sharing octahedra, perovskite corner-sharing octahedra and plumed input files, are freely available from the corresponding author upon request. In future, these codes will be made available in a new release of the PLUMED software. 

\subsection*{Data availability}
The simulation data that support the findings of this study is available from the corresponding author upon request.

\newpage
%\printbibliography

\bibliographystyle{naturemag}

\bibliography{1st_Paper.bib}

\begin{thebibliography}{10}
\expandafter\ifx\csname url\endcsname\relax
  \def\url#1{\texttt{#1}}\fi
\expandafter\ifx\csname urlprefix\endcsname\relax\def\urlprefix{URL }\fi
\providecommand{\bibinfo}[2]{#2}
\providecommand{\eprint}[2][]{\url{#2}}

\bibitem{Gratzel2014}
\bibinfo{author}{Gr{\"{a}}tzel, M.}
\newblock \bibinfo{title}{{The light and shade of perovskite solar cells}}.
\newblock \emph{\bibinfo{journal}{Nat. Mater.}} \textbf{\bibinfo{volume}{13}},
  \bibinfo{pages}{838--842} (\bibinfo{year}{2014}).

\bibitem{Yang2017}
\bibinfo{author}{Yang, W.~S.} \emph{et~al.}
\newblock \bibinfo{title}{{Iodide management in formamidinium-lead-halide-based
  perovskite layers for efficient solar cells.}}
\newblock \emph{\bibinfo{journal}{Science}} \textbf{\bibinfo{volume}{356}},
  \bibinfo{pages}{1376--1379} (\bibinfo{year}{2017}).

\bibitem{Saliba2016a}
\bibinfo{author}{Saliba, M.} \emph{et~al.}
\newblock \bibinfo{title}{{Incorporation of rubidium cations into perovskite
  solar cells improves photovoltaic performance.}}
\newblock \emph{\bibinfo{journal}{Science}} \textbf{\bibinfo{volume}{354}},
  \bibinfo{pages}{206--209} (\bibinfo{year}{2016}).

\bibitem{Tan2014}
\bibinfo{author}{Tan, Z.-K.} \emph{et~al.}
\newblock \bibinfo{title}{{Bright light-emitting diodes based on organometal
  halide perovskite}}.
\newblock \emph{\bibinfo{journal}{Nat. Nanotechnol.}}
  \textbf{\bibinfo{volume}{9}}, \bibinfo{pages}{687--692}
  (\bibinfo{year}{2014}).

\bibitem{Fang2015}
\bibinfo{author}{Fang, Y.}, \bibinfo{author}{Dong, Q.}, \bibinfo{author}{Shao,
  Y.}, \bibinfo{author}{Yuan, Y.} \& \bibinfo{author}{Huang, J.}
\newblock \bibinfo{title}{{Highly narrowband perovskite single-crystal
  photodetectors enabled by surface-charge recombination}}.
\newblock \emph{\bibinfo{journal}{Nat. Photonics}}
  \textbf{\bibinfo{volume}{9}}, \bibinfo{pages}{679--686}
  (\bibinfo{year}{2015}).

\bibitem{Sharenko2016}
\bibinfo{author}{Sharenko, A.} \& \bibinfo{author}{Toney, M.~F.}
\newblock \bibinfo{title}{{Relationships between Lead Halide Perovskite
  Thin-Film Fabrication, Morphology, and Performance in Solar Cells}}.
\newblock \emph{\bibinfo{journal}{J. Am. Chem. Soc.}}
  \textbf{\bibinfo{volume}{138}}, \bibinfo{pages}{463--470}
  (\bibinfo{year}{2016}).

\bibitem{Salim2015}
\bibinfo{author}{Salim, T.} \emph{et~al.}
\newblock \bibinfo{title}{{Perovskite-based solar cells: impact of morphology
  and device architecture on device performance}}.
\newblock \emph{\bibinfo{journal}{J. Mater. Chem. A}}
  \textbf{\bibinfo{volume}{3}}, \bibinfo{pages}{8943--8969}
  (\bibinfo{year}{2015}).

\bibitem{Bi2016}
\bibinfo{author}{Bi, D.} \emph{et~al.}
\newblock \bibinfo{title}{{Efficient luminescent solar cells based on tailored
  mixed-cation perovskites}}.
\newblock \emph{\bibinfo{journal}{Sci. Adv.}} \textbf{\bibinfo{volume}{2}},
  \bibinfo{pages}{e1501170--e1501170} (\bibinfo{year}{2016}).

\bibitem{Snaith2018}
\bibinfo{author}{Snaith, H.~J.}
\newblock \bibinfo{title}{{Present status and future prospects of perovskite
  photovoltaics}}.
\newblock \emph{\bibinfo{journal}{Nat. Mater.}} \textbf{\bibinfo{volume}{17}},
  \bibinfo{pages}{372--376} (\bibinfo{year}{2018}).

\bibitem{Huang2016}
\bibinfo{author}{Huang, H.} \emph{et~al.}
\newblock \bibinfo{title}{{Colloidal lead halide perovskite nanocrystals:
  synthesis, optical properties and applications}}.
\newblock \emph{\bibinfo{journal}{NPG Asia Mater.}}
  \textbf{\bibinfo{volume}{8}}, \bibinfo{pages}{e328--e328}
  (\bibinfo{year}{2016}).

\bibitem{Akkerman2018}
\bibinfo{author}{Akkerman, Q.~A.}, \bibinfo{author}{Rain{\`{o}}, G.},
  \bibinfo{author}{Kovalenko, M.~V.} \& \bibinfo{author}{Manna, L.}
\newblock \bibinfo{title}{{Genesis, challenges and opportunities for colloidal
  lead halide perovskite nanocrystals}}.
\newblock \emph{\bibinfo{journal}{Nat. Mater.}} \bibinfo{pages}{1}
  (\bibinfo{year}{2018}).

\bibitem{Ball2016}
\bibinfo{author}{Ball, J.~M.} \& \bibinfo{author}{Petrozza, A.}
\newblock \bibinfo{title}{{Defects in perovskite-halides and their effects in
  solar cells}}.
\newblock \emph{\bibinfo{journal}{Nat. Energy}} \textbf{\bibinfo{volume}{1}},
  \bibinfo{pages}{16149} (\bibinfo{year}{2016}).

\bibitem{Snaith2014}
\bibinfo{author}{Snaith, H.~J.} \emph{et~al.}
\newblock \bibinfo{title}{{Anomalous Hysteresis in Perovskite Solar Cells}}.
\newblock \emph{\bibinfo{journal}{J. Phys. Chem. Lett.}}
  \textbf{\bibinfo{volume}{5}}, \bibinfo{pages}{1511--1515}
  (\bibinfo{year}{2014}).

\bibitem{Eperon2014}
\bibinfo{title}{{Morphological Control for High Performance, Solution-Processed
  Planar Heterojunction Perovskite Solar Cells}}.
\newblock \emph{\bibinfo{journal}{Adv. Funct. Mater.}}
  \textbf{\bibinfo{volume}{24}}, \bibinfo{pages}{151--157}
  (\bibinfo{year}{2014}).

\bibitem{Shi2015}
\bibinfo{author}{Shi, D.} \emph{et~al.}
\newblock \bibinfo{title}{{Solar cells. Low trap-state density and long carrier
  diffusion in organolead trihalide perovskite single crystals.}}
\newblock \emph{\bibinfo{journal}{Science}} \textbf{\bibinfo{volume}{347}},
  \bibinfo{pages}{519--22} (\bibinfo{year}{2015}).

\bibitem{Burschka2013}
\bibinfo{author}{Burschka, J.} \emph{et~al.}
\newblock \bibinfo{title}{{Sequential deposition as a route to high-performance
  perovskite-sensitized solar cells}}.
\newblock \emph{\bibinfo{journal}{Nature}} \textbf{\bibinfo{volume}{499}},
  \bibinfo{pages}{316--319} (\bibinfo{year}{2013}).

\bibitem{Im2011}
\bibinfo{author}{Im, J.-H.}, \bibinfo{author}{Lee, C.-R.},
  \bibinfo{author}{Lee, J.-W.}, \bibinfo{author}{Park, S.-W.} \&
  \bibinfo{author}{Park, N.-G.}
\newblock \bibinfo{title}{6.5{\%} efficient perovskite quantum-dot-sensitized
  solar cell}.
\newblock \emph{\bibinfo{journal}{Nanoscale}} \textbf{\bibinfo{volume}{3}},
  \bibinfo{pages}{4088} (\bibinfo{year}{2011}).

\bibitem{KangningLiang1998}
\bibinfo{author}{{Kangning Liang}}, \bibinfo{author}{{David B. Mitzi}} \&
  \bibinfo{author}{Prikas, M.~T.}
\newblock \bibinfo{title}{{Synthesis and Characterization of
  Organic−Inorganic Perovskite Thin Films Prepared Using a Versatile Two-Step
  Dipping Technique}}  (\bibinfo{year}{1998}).

\bibitem{Dar2014}
\bibinfo{author}{Dar, M.~I.} \emph{et~al.}
\newblock \bibinfo{title}{{Investigation Regarding the Role of Chloride in
  Organic–Inorganic Halide Perovskites Obtained from Chloride Containing
  Precursors}}.
\newblock \emph{\bibinfo{journal}{Nano Lett.}} \textbf{\bibinfo{volume}{14}},
  \bibinfo{pages}{6991--6996} (\bibinfo{year}{2014}).

\bibitem{Jeon2014}
\bibinfo{author}{Jeon, N.~J.} \emph{et~al.}
\newblock \bibinfo{title}{{Solvent engineering for high-performance
  inorganic–organic hybrid perovskite solar cells}}.
\newblock \emph{\bibinfo{journal}{Nat. Mater.}} \textbf{\bibinfo{volume}{13}},
  \bibinfo{pages}{897--903} (\bibinfo{year}{2014}).

\bibitem{Dualeh2014}
\bibinfo{author}{Dualeh, A.} \emph{et~al.}
\newblock \bibinfo{title}{{Effect of Annealing Temperature on Film Morphology
  of Organic-Inorganic Hybrid Pervoskite Solid-State Solar Cells}}.
\newblock \emph{\bibinfo{journal}{Adv. Funct. Mater.}}
  \textbf{\bibinfo{volume}{24}}, \bibinfo{pages}{3250--3258}
  (\bibinfo{year}{2014}).

\bibitem{Ummadisingu2017}
\bibinfo{author}{Ummadisingu, A.} \emph{et~al.}
\newblock \bibinfo{title}{{The effect of illumination on the formation of metal
  halide perovskite films}}.
\newblock \emph{\bibinfo{journal}{Nature}} \textbf{\bibinfo{volume}{545}},
  \bibinfo{pages}{208--212} (\bibinfo{year}{2017}).

\bibitem{Tsai2018}
\bibinfo{author}{Tsai, H.} \emph{et~al.}
\newblock \bibinfo{title}{{Light-induced lattice expansion leads to
  high-efficiency perovskite solar cells.}}
\newblock \emph{\bibinfo{journal}{Science}} \textbf{\bibinfo{volume}{360}},
  \bibinfo{pages}{67--70} (\bibinfo{year}{2018}).

\bibitem{Moore2015}
\bibinfo{author}{Moore, D.~T.} \emph{et~al.}
\newblock \bibinfo{title}{{Crystallization Kinetics of Organic–Inorganic
  Trihalide Perovskites and the Role of the Lead Anion in Crystal Growth}}.
\newblock \emph{\bibinfo{journal}{J. Am. Chem. Soc.}}
  \textbf{\bibinfo{volume}{137}}, \bibinfo{pages}{2350--2358}
  (\bibinfo{year}{2015}).

\bibitem{Manser2015}
\bibinfo{author}{Manser, J.~S.}, \bibinfo{author}{Reid, B.} \&
  \bibinfo{author}{Kamat, P.~V.}
\newblock \bibinfo{title}{{Evolution of Organic–Inorganic Lead Halide
  Perovskite from Solid-State Iodoplumbate Complexes}}.
\newblock \emph{\bibinfo{journal}{J. Phys. Chem. C}}
  \textbf{\bibinfo{volume}{119}}, \bibinfo{pages}{17065--17073}
  (\bibinfo{year}{2015}).

\bibitem{Rahimnejad2016}
\bibinfo{author}{Rahimnejad, S.}, \bibinfo{author}{Kovalenko, A.},
  \bibinfo{author}{For{\'{e}}s, S.~M.}, \bibinfo{author}{Aranda, C.} \&
  \bibinfo{author}{Guerrero, A.}
\newblock \bibinfo{title}{{Coordination Chemistry Dictates the Structural
  Defects in Lead Halide Perovskites}}.
\newblock \emph{\bibinfo{journal}{ChemPhysChem}} \textbf{\bibinfo{volume}{17}},
  \bibinfo{pages}{2795--2798} (\bibinfo{year}{2016}).

\bibitem{Stewart2016}
\bibinfo{author}{Stewart, R.~J.}, \bibinfo{author}{Grieco, C.},
  \bibinfo{author}{Larsen, A.~V.}, \bibinfo{author}{Doucette, G.~S.} \&
  \bibinfo{author}{Asbury, J.~B.}
\newblock \bibinfo{title}{{Molecular Origins of Defects in Organohalide
  Perovskites and Their Influence on Charge Carrier Dynamics}}.
\newblock \emph{\bibinfo{journal}{J. Phys. Chem. C}}
  \textbf{\bibinfo{volume}{120}}, \bibinfo{pages}{12392--12402}
  (\bibinfo{year}{2016}).

\bibitem{Yan2015}
\bibinfo{author}{Yan, K.} \emph{et~al.}
\newblock \bibinfo{title}{{Hybrid Halide Perovskite Solar Cell Precursors:
  Colloidal Chemistry and Coordination Engineering behind Device Processing for
  High Efficiency}}.
\newblock \emph{\bibinfo{journal}{J. Am. Chem. Soc.}}
  \textbf{\bibinfo{volume}{137}}, \bibinfo{pages}{4460--4468}
  (\bibinfo{year}{2015}).

\bibitem{McMeekin2017}
\bibinfo{author}{McMeekin, D.~P.} \emph{et~al.}
\newblock \bibinfo{title}{{Crystallization Kinetics and Morphology Control of
  Formamidinium-Cesium Mixed-Cation Lead Mixed-Halide Perovskite via Tunability
  of the Colloidal Precursor Solution}}.
\newblock \emph{\bibinfo{journal}{Adv. Mater.}} \textbf{\bibinfo{volume}{29}},
  \bibinfo{pages}{1607039} (\bibinfo{year}{2017}).

\bibitem{Hu2017}
\bibinfo{author}{Hu, Q.} \emph{et~al.}
\newblock \bibinfo{title}{{In situ dynamic observations of perovskite
  crystallisation and microstructure evolution intermediated from
  [PbI$_6$]$^{4-}$ cage nanoparticles}}.
\newblock \emph{\bibinfo{journal}{Nat. Commun.}} \textbf{\bibinfo{volume}{8}},
  \bibinfo{pages}{15688} (\bibinfo{year}{2017}).

\bibitem{Ummadisingue1}
\bibinfo{author}{Ummadisingu, A.} \& \bibinfo{author}{Gr{\"a}tzel, M.}
\newblock \bibinfo{title}{Revealing the detailed path of sequential deposition
  for metal halide perovskite formation}.
\newblock \emph{\bibinfo{journal}{Science Advances}}
  \textbf{\bibinfo{volume}{4}} (\bibinfo{year}{2018}).

\bibitem{Salvalaglio2015}
\bibinfo{author}{Salvalaglio, M.}, \bibinfo{author}{Mazzotti, M.} \&
  \bibinfo{author}{Parrinello, M.}
\newblock \bibinfo{title}{{Urea homogeneous nucleation mechanism is solvent
  dependent}}.
\newblock \emph{\bibinfo{journal}{Faraday Discuss.}}
  \textbf{\bibinfo{volume}{179}}, \bibinfo{pages}{291--307}
  (\bibinfo{year}{2015}).

\bibitem{Salvalaglio2015a}
\bibinfo{author}{Salvalaglio, M.}, \bibinfo{author}{Perego, C.},
  \bibinfo{author}{Giberti, F.}, \bibinfo{author}{Mazzotti, M.} \&
  \bibinfo{author}{Parrinello, M.}
\newblock \bibinfo{title}{{Molecular-dynamics simulations of urea nucleation
  from aqueous solution.}}
\newblock \emph{\bibinfo{journal}{Proc. Natl. Acad. Sci. U. S. A.}}
  \textbf{\bibinfo{volume}{112}}, \bibinfo{pages}{E6--14}
  (\bibinfo{year}{2015}).

\bibitem{Laio2002}
\bibinfo{author}{Laio, A.} \& \bibinfo{author}{Parrinello, M.}
\newblock \bibinfo{title}{{Escaping free-energy minima.}}
\newblock \emph{\bibinfo{journal}{Proc. Natl. Acad. Sci. U. S. A.}}
  \textbf{\bibinfo{volume}{99}}, \bibinfo{pages}{12562--6}
  (\bibinfo{year}{2002}).

\bibitem{Barducci2008}
\bibinfo{author}{Barducci, A.}, \bibinfo{author}{Bussi, G.} \&
  \bibinfo{author}{Parrinello, M.}
\newblock \bibinfo{title}{{Well-Tempered Metadynamics: A Smoothly Converging
  and Tunable Free-Energy Method}}.
\newblock \emph{\bibinfo{journal}{Phys. Rev. Lett.}}
  \textbf{\bibinfo{volume}{100}}, \bibinfo{pages}{020603}
  (\bibinfo{year}{2008}).

\bibitem{Saidaminov2015}
\bibinfo{title}{{High-quality bulk hybrid perovskite single crystals within
  minutes by inverse temperature crystallization}}.
\newblock \emph{\bibinfo{journal}{Nat. Commun.}} \textbf{\bibinfo{volume}{6}},
  \bibinfo{pages}{7586} (\bibinfo{year}{2015}).

\bibitem{Heo2013}
\bibinfo{author}{Heo, J.~H.} \emph{et~al.}
\newblock \bibinfo{title}{{Efficient inorganic–organic hybrid heterojunction
  solar cells containing perovskite compound and polymeric hole conductors}}.
\newblock \emph{\bibinfo{journal}{Nat. Photonics}}
  \textbf{\bibinfo{volume}{7}}, \bibinfo{pages}{486--491}
  (\bibinfo{year}{2013}).

\bibitem{Kadro2015}
\bibinfo{author}{Kadro, J.~M.}, \bibinfo{author}{Nonomura, K.},
  \bibinfo{author}{Gachet, D.}, \bibinfo{author}{Gr{\"{a}}tzel, M.} \&
  \bibinfo{author}{Hagfeldt, A.}
\newblock \bibinfo{title}{{Facile route to freestanding CH3NH3PbI3 crystals
  using inverse solubility}}.
\newblock \emph{\bibinfo{journal}{Sci. Rep.}} \textbf{\bibinfo{volume}{5}},
  \bibinfo{pages}{11654} (\bibinfo{year}{2015}).

\bibitem{Zhang2017a}
\bibinfo{author}{Zhang, F.} \emph{et~al.}
\newblock \bibinfo{title}{{Colloidal Synthesis of Air-Stable
  CH$_3$NH$_3$PbI$_3$ Quantum Dots by Gaining Chemical Insight into the Solvent
  Effects}}.
\newblock \emph{\bibinfo{journal}{Chem. Mater.}} \textbf{\bibinfo{volume}{29}},
  \bibinfo{pages}{3793--3799} (\bibinfo{year}{2017}).

\bibitem{Humphrey1996}
\bibinfo{author}{Humphrey, W.}, \bibinfo{author}{Dalke, A.} \&
  \bibinfo{author}{Schulten, K.}
\newblock \bibinfo{title}{{VMD: Visual molecular dynamics}}.
\newblock \emph{\bibinfo{journal}{J. Mol. Graph.}}
  \textbf{\bibinfo{volume}{14}}, \bibinfo{pages}{33--38}
  (\bibinfo{year}{1996}).

\bibitem{Ahn2015}
\bibinfo{author}{Ahn, N.} \emph{et~al.}
\newblock \bibinfo{title}{{Highly Reproducible Perovskite Solar Cells with
  Average Efficiency of 18.3{\%} and Best Efficiency of 19.7{\%} Fabricated via
  Lewis Base Adduct of Lead(II) Iodide}}.
\newblock \emph{\bibinfo{journal}{J. Am. Chem. Soc.}}
  \textbf{\bibinfo{volume}{137}}, \bibinfo{pages}{8696--8699}
  (\bibinfo{year}{2015}).

\bibitem{Matsui2017}
\bibinfo{author}{Matsui, T.}, \bibinfo{author}{Seo, J.-Y.},
  \bibinfo{author}{Saliba, M.}, \bibinfo{author}{Zakeeruddin, S.~M.} \&
  \bibinfo{author}{Gr{\"{a}}tzel, M.}
\newblock \bibinfo{title}{{Room-Temperature Formation of Highly Crystalline
  Multication Perovskites for Efficient, Low-Cost Solar Cells}}.
\newblock \emph{\bibinfo{journal}{Adv. Mater.}} \textbf{\bibinfo{volume}{29}},
  \bibinfo{pages}{1606258} (\bibinfo{year}{2017}).

\bibitem{Yi2016}
\bibinfo{author}{Yi, C.} \emph{et~al.}
\newblock \bibinfo{title}{{Entropic stabilization of mixed A-cation ABX$_3$
  metal halide perovskites for high performance perovskite solar cells}}.
\newblock \emph{\bibinfo{journal}{Energy Environ. Sci.}}
  \textbf{\bibinfo{volume}{9}}, \bibinfo{pages}{656--662}
  (\bibinfo{year}{2016}).

\bibitem{Mattoni2015}
\bibinfo{author}{Mattoni, A.}, \bibinfo{author}{Filippetti, A.},
  \bibinfo{author}{Saba, M.~I.} \& \bibinfo{author}{Delugas, P.}
\newblock \bibinfo{title}{{Methylammonium Rotational Dynamics in Lead Halide
  Perovskite by Classical Molecular Dynamics: The Role of Temperature}}.
\newblock \emph{\bibinfo{journal}{J. Phys. Chem. C}}
  \textbf{\bibinfo{volume}{119}}, \bibinfo{pages}{17421--17428}
  (\bibinfo{year}{2015}).

\bibitem{Gutierrez-Sevillano2015a}
\bibinfo{author}{Gutierrez-Sevillano, J.~J.}, \bibinfo{author}{Ahmad, S.},
  \bibinfo{author}{Calero, S.} \& \bibinfo{author}{Anta, J.~A.}
\newblock \bibinfo{title}{{Molecular dynamics simulations of organohalide
  perovskite precursors: solvent effects in the formation of perovskite solar
  cells}}.
\newblock \emph{\bibinfo{journal}{Phys. Chem. Chem. Phys.}}
  \textbf{\bibinfo{volume}{17}}, \bibinfo{pages}{22770--22777}
  (\bibinfo{year}{2015}).

\bibitem{Ryckaert1977}
\bibinfo{author}{Ryckaert, J.-P.}, \bibinfo{author}{Ciccotti, G.} \&
  \bibinfo{author}{Berendsen, H.~J.}
\newblock \bibinfo{title}{{Numerical integration of the cartesian equations of
  motion of a system with constraints: molecular dynamics of n-alkanes}}.
\newblock \emph{\bibinfo{journal}{J. Comput. Phys.}}
  \textbf{\bibinfo{volume}{23}}, \bibinfo{pages}{327--341}
  (\bibinfo{year}{1977}).

\bibitem{Plimpton1995}
\bibinfo{author}{Plimpton, S.}
\newblock \bibinfo{title}{{Fast Parallel Algorithms for Short-Range Molecular
  Dynamics}}.
\newblock \emph{\bibinfo{journal}{J. Comput. Phys.}}
  \textbf{\bibinfo{volume}{117}}, \bibinfo{pages}{1--19}
  (\bibinfo{year}{1995}).

\bibitem{Bussi2007}
\bibinfo{author}{Bussi, G.}, \bibinfo{author}{Donadio, D.} \&
  \bibinfo{author}{Parrinello, M.}
\newblock \bibinfo{title}{{Canonical sampling through velocity rescaling}}.
\newblock \emph{\bibinfo{journal}{J. Chem. Phys.}}
  \textbf{\bibinfo{volume}{126}}, \bibinfo{pages}{014101}
  (\bibinfo{year}{2007}).

\bibitem{Parrinello1981}
\bibinfo{author}{Parrinello, M.} \& \bibinfo{author}{Rahman, A.}
\newblock \bibinfo{title}{{Polymorphic transitions in single crystals: A new
  molecular dynamics method}}.
\newblock \emph{\bibinfo{journal}{J. Appl. Phys.}}
  \textbf{\bibinfo{volume}{52}}, \bibinfo{pages}{7182--7190}
  (\bibinfo{year}{1981}).

\bibitem{Tribello2014}
\bibinfo{author}{Tribello, G.~A.}, \bibinfo{author}{Bonomi, M.},
  \bibinfo{author}{Branduardi, D.}, \bibinfo{author}{Camilloni, C.} \&
  \bibinfo{author}{Bussi, G.}
\newblock \bibinfo{title}{{PLUMED 2: New feathers for an old bird}}.
\newblock \emph{\bibinfo{journal}{Comput. Phys. Commun.}}
  \textbf{\bibinfo{volume}{185}}, \bibinfo{pages}{604--613}
  (\bibinfo{year}{2014}).

\end{thebibliography}


\begin{thebibliography}{1}
\expandafter\ifx\csname url\endcsname\relax
  \def\url#1{\texttt{#1}}\fi
\expandafter\ifx\csname urlprefix\endcsname\relax\def\urlprefix{URL }\fi
\providecommand{\bibinfo}[2]{#2}
\providecommand{\eprint}[2][]{\url{#2}}

\bibitem{Tribello2014}
\bibinfo{author}{Tribello, G.~A.}, \bibinfo{author}{Bonomi, M.},
  \bibinfo{author}{Branduardi, D.}, \bibinfo{author}{Camilloni, C.} \&
  \bibinfo{author}{Bussi, G.}
\newblock \bibinfo{title}{{PLUMED 2: New feathers for an old bird}}.
\newblock \emph{\bibinfo{journal}{Comput. Phys. Commun.}}
  \textbf{\bibinfo{volume}{185}}, \bibinfo{pages}{604--613}
  (\bibinfo{year}{2014}).

\bibitem{Tribello2017}
\bibinfo{author}{Tribello, G.~A.}, \bibinfo{author}{Giberti, F.},
  \bibinfo{author}{Sosso, G.~C.}, \bibinfo{author}{Salvalaglio, M.} \&
  \bibinfo{author}{Parrinello, M.}
\newblock \bibinfo{title}{{Analyzing and Driving Cluster Formation in Atomistic
  Simulations}}.
\newblock \emph{\bibinfo{journal}{J. Chem. Theory Comput.}}
  \textbf{\bibinfo{volume}{13}}, \bibinfo{pages}{1317--1327}
  (\bibinfo{year}{2017}).

\bibitem{Humphrey1996}
\bibinfo{author}{Humphrey, W.}, \bibinfo{author}{Dalke, A.} \&
  \bibinfo{author}{Schulten, K.}
\newblock \bibinfo{title}{{VMD: Visual molecular dynamics}}.
\newblock \emph{\bibinfo{journal}{J. Mol. Graph.}}
  \textbf{\bibinfo{volume}{14}}, \bibinfo{pages}{33--38}
  (\bibinfo{year}{1996}).

\end{thebibliography}

\newpage
\section*{Acknowledgements}
P.A. thanks Michele Ceriotti, Omar Valsson, Claudio Perego and Simone Meloni for useful discussions. P.A. thanks in particular to Michele Ceriotti for highly insightful conversation and criticism of the work. P.A. is greately thankful to Claudio Perego and Omar Valsson for their support, encouragement and precious advice. M.P. acknowledge funding from European Union Grant No. ERC-2014-AdG-670227/VARMET. This research is funded by Swiss National Science Foundation through the NCCR MUST and individual grant 200020-165863. The computational time for this work was provided by the Swiss National Supercomputing Center (CSCS) under project s672. All calculations were performed in CSCS clusters Piz Daint.

\section*{Author information}

\subsection*{Affiliations}
%Paramvir Ahlawat$^a$, Pablo Piaggi$^{c,d}$, Michele Parrinello$^{b,c}$, Ursula Roethlisberger$^a$
Laboratory of Computational Chemistry and Biochemistry, Ecole Polytechnique Fédérale de Lausanne, CH-1015 Lausanne, Switzerland; \\
Paramvir Ahlawat \& Ursula Rothlisberger
\\
\\
Department of Chemistry and Applied Biosciences, Eidgenossische Technische Hochschule (ETH) Zurich c/o Università della Svizzera Italiana Campus, 6900 Lugano, Switzerland;\\ 
Michele Parrinello
\\
\\
Facoltà di Informatica, Instituto di Scienze Computationali, and National Center for Computational Design and Discovery of Novel Materials (MARVEL), Università della Svizzera Italiana, 6900 Lugano, Switzerland;\\
Pablo Piaggi \& Michele Parrinello
\\
\\
Theory and Simulation of Materials, École Polytechnique Fédérale de Lausanne, c/o Università della Svizzera Italiana Campus, 6900 Lugano, Switzerland; \\
Pablo Piaggi
\\
\\
Laboratory of Photonics and Interfaces, Ecole polytechnique Fédérale (EPFL) de Lausanne, Lausanne, Switzerland \\
Michael Graetzel

\subsection*{Contributions}
M.G., M.P. and U.R. conceived the research and contributed to the interpretation of the results. P.A. perfomed all the simulations. P.A. and P.P. developed new analytical tools and analyzed the data. All authors discussed the results and wrote the manuscript.

\subsection*{Competing interests}
The authors declare no competing financial interests.
\subsection*{Corresponding author}
Correspondence to Ursula Rothlisberger

\end{document}

%\linenumbers
\maketitle
\section*{Clustering at the first stage}
In order to quantify the number and type of the clusters of Pb\SP{2+} and I\SP{-} at the first stage of nucleation in our simulations, we have performed a clustering analysis with the depth first search (DFS) algorithm. Further details on using this method and the implementation in PLUMED 2.4\cite{Tribello2014} are provided in reference \cite{Tribello2017}. The following plumed input is used for clustering. \\ 

%Plumed input:  \\
%\\
%{\color{blue}\# An example for the simulation-(A)} \\
%{\color{green}Pb: GROUP} {\color{orange}ATOMS=}{\color{red}1-200}  \\
%{\color{green}I: GROUP} {\color{orange}ATOMS=}{\color{red}201-800} \\
%{\color{green}cPbI: COORDINATIONNUMBER} {\color{orange} SPECIESA=}{\color{red}Pb} {\color{orange}SPECIESB=}{\color{red}I} {\color{orange}SWITCH=}{\color{red}\{CUBIC D\_0=0.40 D\_MAX=0.41\}} \\
%{\color{green}cPbI3: MFILTER\_MORE} {\color{orange} DATA=}{\color{red}cPbI} {\color{orange}SWITCH=}{\color{red}\{CUBIC D\_0=2.99 D\_MAX=3.00\}}\\
%{\color{green}cmPbI3: COORDINATIONNUMBER} {\color{orange} SPECIES=}{\color{red}cPbI3} {\color{orange}SWITCH=}{\color{red}\{CUBIC D\_0=0.70 D\_MAX=0.71\}} \\
%{\color{green}cmPbI34: MFILTER\_MORE} {\color{orange} DATA=}{\color{red}cmPbI3} {\color{orange}SWITCH=}{\color{red}\{CUBIC D\_0=3.99 D\_MAX=4.00\}}\\
%{\color{green}cm: CONTACT\_MATRIX} {\color{orange} ATOMS=}{\color{red}cmPbI34} {\color{orange}SWITCH=}{\color{red}\{CUBIC D\_0=0.70 D\_MAX=0.71\}}\\
%{\color{green}dsf: DFSCLUSTERING} {\color{orange} MATRIX=} {\color{red}cm}\\
%{\color{green}nat1: CLUSTER\_NATOMS} {\color{orange} CLUSTERS=} {\color{red}dsf} {\color{orange} CLUSTER=} {\color{red}1} \\
%{\color{green}PRINT:} {\color{orange} ARG=} {\color{red}nat1} {\color{orange} FILE=} {\color{red}COLVAR\_cluster}\\
\begin{tcbdoublebox}[title={Plumed input file:}]
\begin{figure}[H]
 \centering 
 \includegraphics[width=150mm,scale=1]{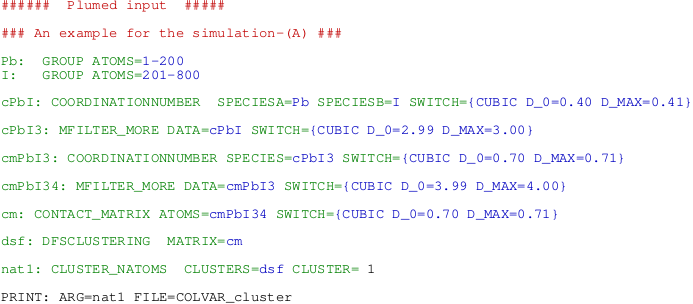}
\end{figure}
\end{tcbdoublebox}

%\includepdf[pages=-, scale=0.9, pagecommand={}]{1st_stage_cluster_plumed.pdf}

%I:  GROUP ATOMS=201-800
%cPbI: COORDINATIONNUMBER SPECIESA=Pb SPECIESB=I SWITCH={CUBIC D_0=0.40 D_MAX=0.41}
%cPbI3: MFILTER_MORE DATA=cPbI SWITCH={CUBIC D_0=2.99 D_MAX=3.00} 
%cmPbI3: COORDINATIONNUMBER SPECIES=cPbI3 SWITCH={CUBIC D_0=0.70 D_MAX=0.71}
%cmPbI34: MFILTER_MORE DATA=cmPbI3 SWITCH={CUBIC D_0=3.99 D_MAX=4.0} 
%cm: CONTACT_MATRIX ATOMS=cmPbI34 SWITCH={CUBIC D_0=0.70 D_MAX=0.71}
%dfs: DFSCLUSTERING MATRIX=cm
%nat1: CLUSTER_NATOMS CLUSTERS=dfs CLUSTER=1
%PRINT  ARG=nat1 FILE=COLVAR_cluster

In the first two lines, the number of Pb\SP{2+} ions (cPbI3) that are coordinated by at least three I\SP{-} ions within a radial cut-off of 0.40 nm are calculated. Then the coordination number of these Pb\SP{2+} ions (cPbI3) with each other is calculated and only those Pb\SP{2+} ions (cmPbI34) are selected which have a coordination number of at least four between them with a radial cutoff of 0.70 nm. An adjacency matrix is built between cmPbI34 within a cutoff distance of 0.70 nm and the corresponding cluster is calculated with the DFS algorithm. 

\section*{Edge-sharing octahedra}
%Since the positions of Pb\SP{2+} and I\SP{-} ions are fluctuating therefore we decide to take this larger cut-off as compared to nearest distance between Pb-I in MAPI perovskite structure. We only select the nearest six I\SP{-} to mark them as [PbI\SB{6}] octahedra.

\begin{figure}[H]
\centering
  \includegraphics[width=150mm,scale=1]{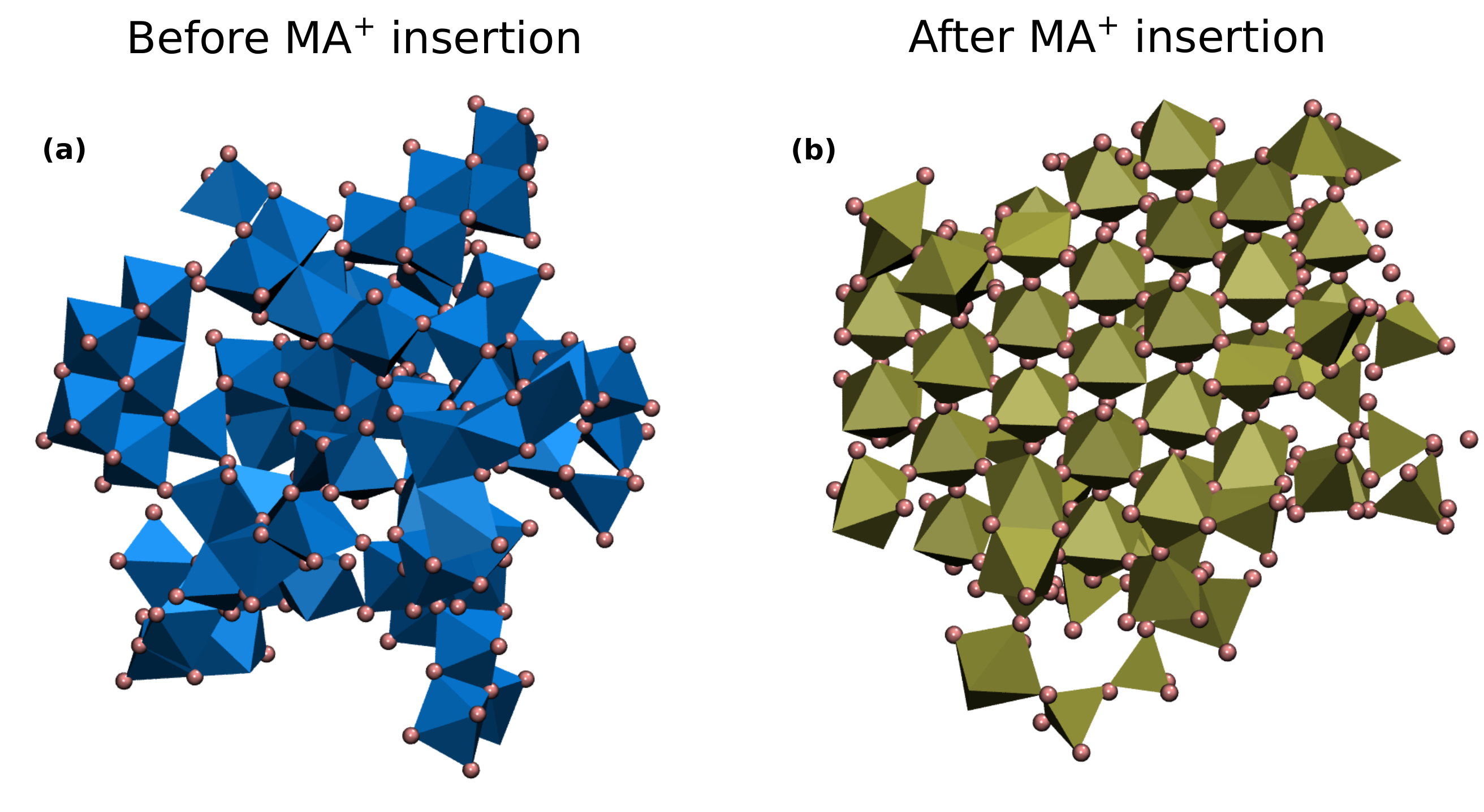}
  \caption{Left panel (a) shows the configuration of edge-sharing octahedra at $\sim$161ns. Pb-I geometry is shown with blue octahedra with Pb\SP{2+} in the center and I\SP{-} on the corners. I\SP{-} is shown in pink spheres. Right panel (b) shows the configuration of the same species at $\sim$327ns. Pb-I geometry is shown with golden octahedra with Pb\SP{2+} in the center and I\SP{-} on the corners. All the images of these snapshots are generated with VMD-1.9.2 \cite{Humphrey1996}}
  \label{fig:edge}
\end{figure}

\begin{figure}[H]
\centering
  \includegraphics[width=100mm,scale=1]{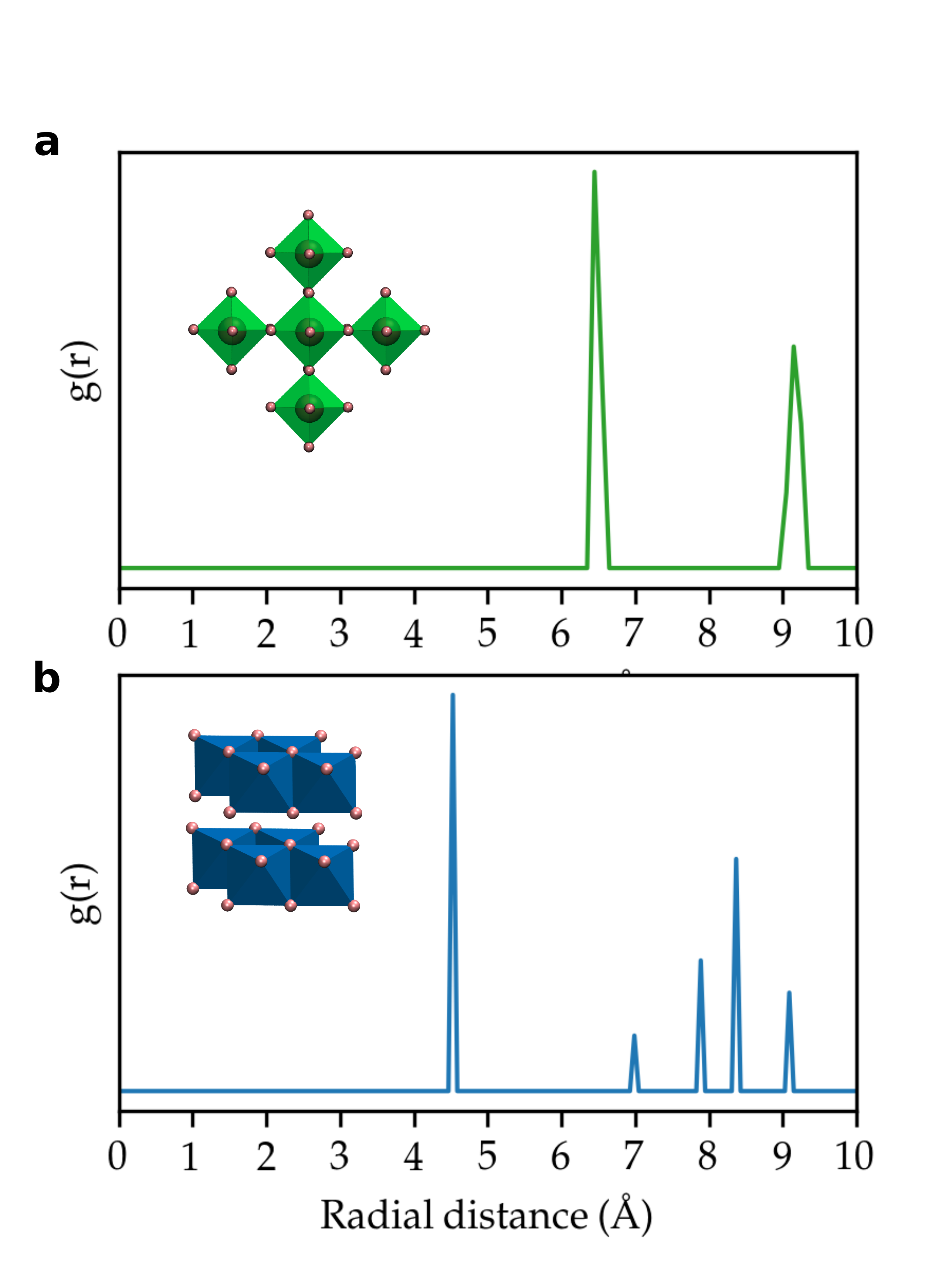}
    \caption{Top panel is the g(r) of Pb\SP{2+}-Pb\SP{2+} in MAPI cubic perovskite and bottom panel is the g(r) of Pb\SP{2+}-Pb\SP{2+} in crystalline PbI\SB{2}}.
  \label{fig:gor}  
\end{figure}

\begin{figure}[H]
\centering
  \includegraphics[width=150mm,scale=1]{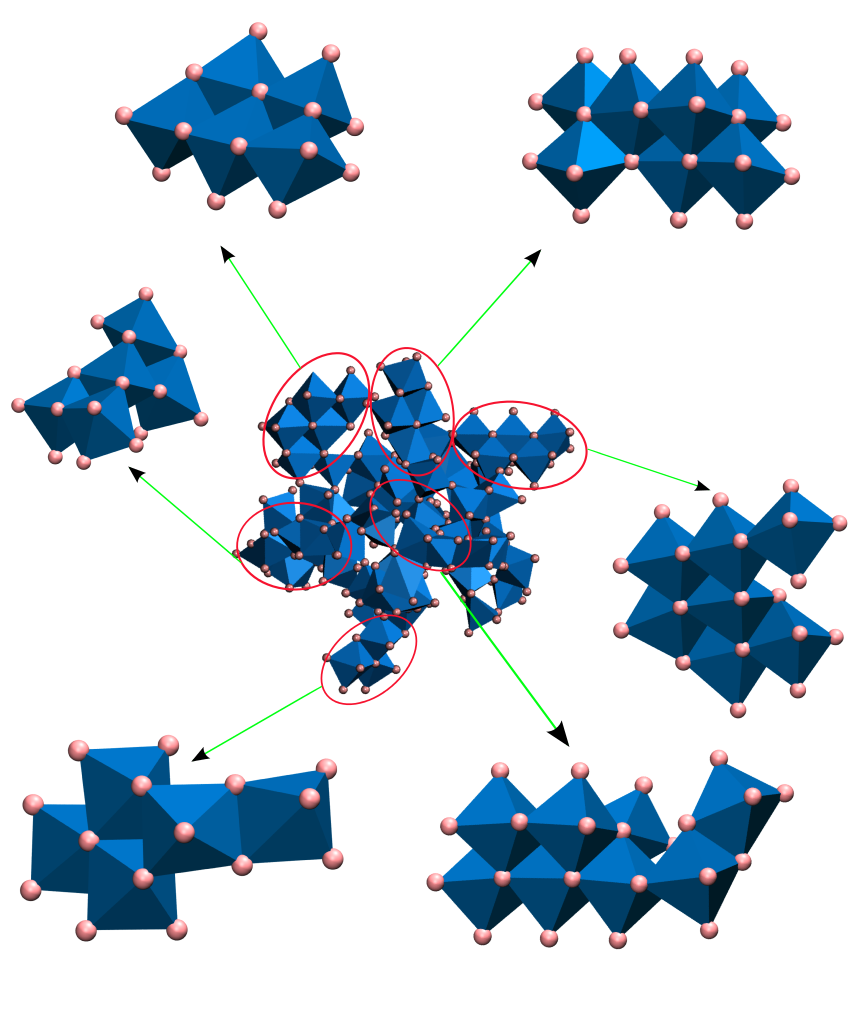}
  \caption{The picture in the middle represents the clusters of edge-sharing octahedra in our simulation (A) at $\sim$ 161 ns. Pb-I complexes are shown as blue octahedra with Pb\SP{2+} in the middle and I\SP{-} on corners. I\SP{-} is shown as pink spheres. Individually arranged structures within edge-sharing clusters are displayed individually. All the images of these snapshots are generated with VMD-1.9.2 \cite{Humphrey1996}}
  \label{fig:edge_sharing_cluster}
\end{figure}

\begin{figure}[H]
\centering
  \includegraphics[width=150mm,scale=1]{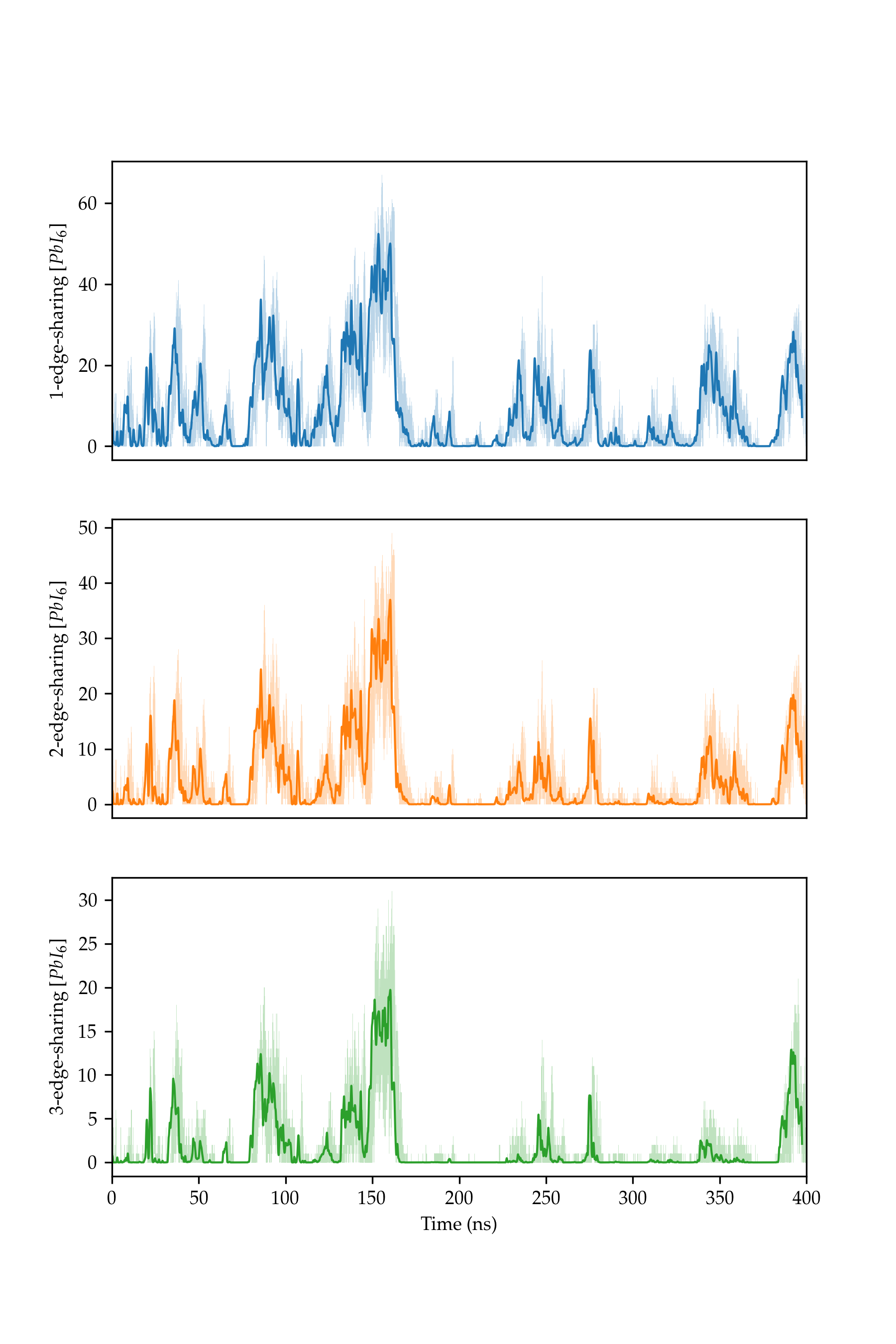}
  \caption{Number of edge-sharing octahedra as a function of simulation time in simulation (A)}
  \label{fig:edge_sharing}
\end{figure}

To calculate the number of edge-sharing octahedra, we identify the Pb\SP{2+} ions that have a six fold coordination with I\SP{-} in a radial cutoff of 0.40 nm as described in the Methods section in the equations for $S_{PbI_6}$. Since the positions of Pb\SP{2+} and I\SP{-} ions are fluctuating continuously in our simulations, we decide to take a larger cutoff as compared to the nearest distance between Pb\SP{2+} and I\SP{-} in the MAPI perovskite structure. However, our algorithm only selects the nearest six I\SP{-} to Pb\SP{2+} and assigns them as [PbI\SB{6}]\SP{4-} octahedra. Then we quantify them on the basis of number of I\SP{-} shared between two neighbouring [PbI\SB{6}]\SP{4-} in a distance of 0.50nm between them. This distance cutoff is chosen according to the first coordination sphere of Pb\SP{2+} atoms in the crystalline P-3\textit{m}1 structure of PbI\SB{2}. The partial radial distribution function (g(r)) of Pb\SP{2+}-Pb\SP{2+} ions in the crystalline cubic phase of MAPI and the P-3\textit{m}1 phase of PbI\SB{2} is shown in Fig. \ref{fig:gor}. A [PbI\SB{6}]\SP{4-} octahedra is identified as an edge-sharing octahedra if the indices of any two of six I\SP{-} that take part in the formation of this [PbI\SB{6}]\SP{4-} octahedra are the same as any two out of six I\SP{-} of neighbouring [PbI\SB{6}]\SP{4-} octahedra. Edge-sharing [PbI\SB{6}]\SP{4-} octahedra are classified based on the number of edges shared between them. We call a [PbI\SB{6}]\SP{4-} octahedron as 1-edge-sharing if at least one-edge is formed between them, 2-edge-sharing if a [PbI\SB{6}]\SP{4-} octahedron shares at least two edges with two different [PbI\SB{6}]\SP{4-} octahedra, 3-edge-sharing shares at least three edges with three different [PbI\SB{6}]\SP{4-} octahedra. In Fig. \ref{fig:edge_sharing} we show the time-evolution of the number of edge-sharing octahedra in simulation (A). It can be noticed that all types of edge-sharing octahedra reach maxima at the second stage $\sim$ 161ns. At this time, we observe aggregates of edge-sharing octahedra as displayed in Fig. \ref{fig:edge_sharing_cluster}. Since it is dificult to see the arranged PbI\SB{2}-like structures in this cluster, therefore for clarity, in Fig. \ref{fig:edge_sharing_cluster} we also display the individual regions of the cluster. These smaller clusters of edge-sharing octahedra show remarkable similarity with the unit cell of PbI\SB{2}. We also track the evolution of these edge-sharing octahedra in our simulation. From Fig. \ref{fig:edge}, one can see that almost all of the edge-sharing octahedra in simulation (A) get converted into a perovskite structure at $\sim$ 327ns. 

%\begin{figure}[H]
%  \centering
%   \includegraphics[width=120mm,scale=0.8]{../Publish_scripts/Nucleation_pics_all/cluster_edge_sharing.png}
%  \caption{Clustering of edge-sharing PbI octahedra}
%  \label{fig:edge_cluster}
%\end{figure}

%\begin{figure}[H]
% \includegraphics[width=80mm,scale=0.8]{../Publish_scripts/Nucleation_pics_all/PbI46.png}
%  \caption{PbI\SB{4} and PbI\SB{6}}
%  \label{fig:PbI46}
%\end{figure}

\section*{Perovskite corner-sharing octahedra}

%\begin{figure}[H]
%  \includegraphics[width=50mm,scale=0.5]{../Publish_scripts/Nucleation_pics_all/Histo_PbIPb.png}
%  \caption{Pb-I-Pb angle histogram in MAPI crystal}
%  \label{fig:histo_PbI_angle}
%\end{figure}

%\begin{figure}[H]
%\centering
%\begin{subfigure}{.2\textwidth}
%  \centering
%    \caption{}
%   \includegraphics[width=30mm,scale=1]{../../Publish_scripts/Nucleation_pics_all/corner2.png}
%  \label{fig:corner-sharing1}
%\end{subfigure}%
%\begin{subfigure}{.2\textwidth}
%  \centering
%   \caption{}
%    \includegraphics[width=30mm,scale=1]{../../Publish_scripts/Nucleation_pics_all/corner3.png}
%  \label{fig:corner-sharing2}
%\end{subfigure}
%\begin{subfigure}{.55\textwidth}
%  \centering
%   \caption{}
%\includegraphics[width=50mm,scale=1]{../../Publish_scripts/Nucleation_pics_all/histo_PbIPb.png}
%  \label{fig:histo_PbI_angle}    
%\end{subfigure}
\begin{figure}[H]
\centering
  \includegraphics[width=120mm,scale=1]{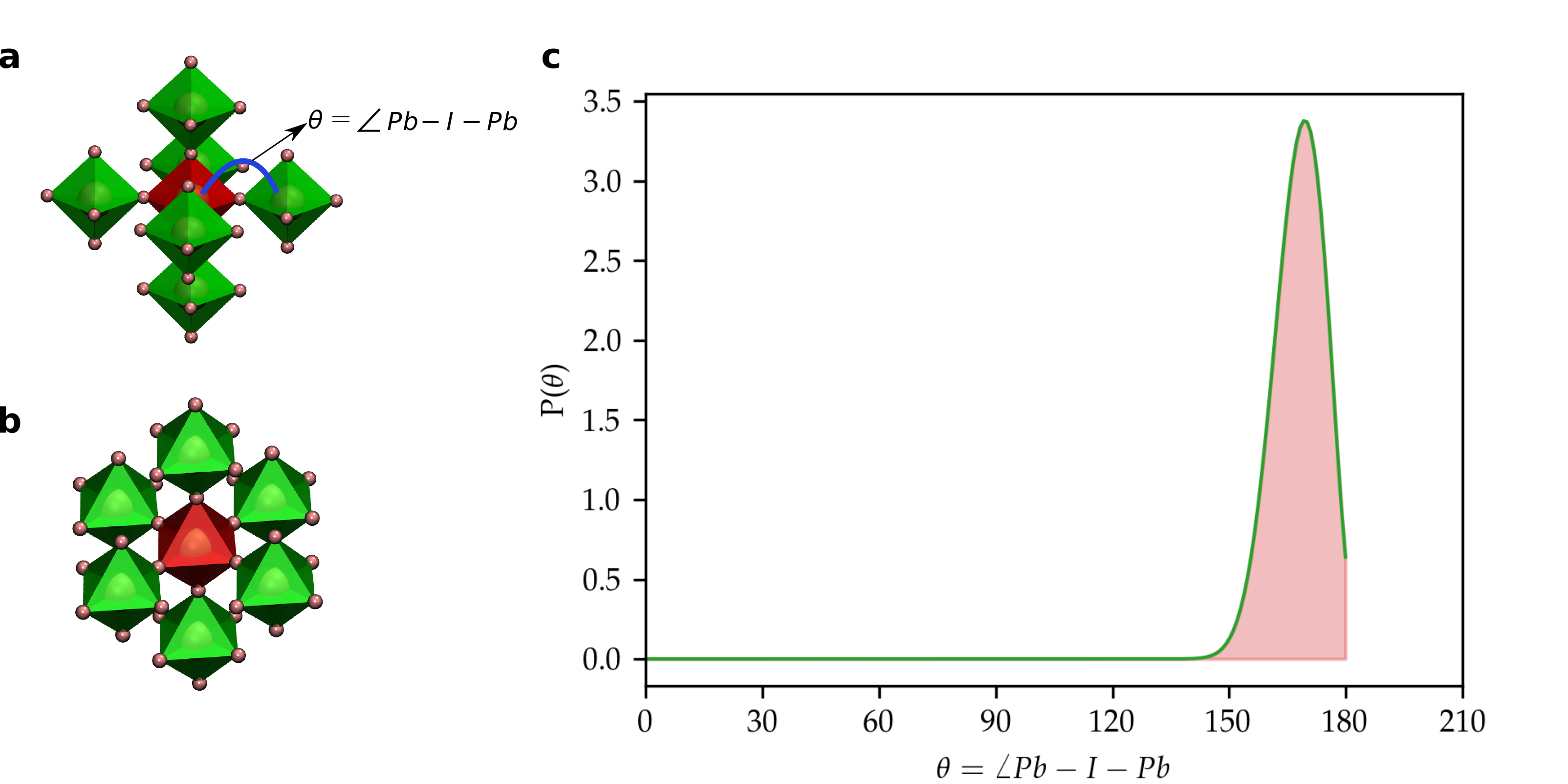}
  \caption{\textbf{Perovskite corner-sharing octahedra}.The two snapshots (a) and (b) on the left shows the configurations of perovskite corner-sharing octahedra. The picture (c) on the right is the histogram of Pb-I-Pb anlges in the cubic perovskite structure of MAPI.}
  \label{fig:Perovskite-corner-sharing}
\end{figure}

Perovskite corner-sharing octahedra are calculated in the following steps: 

1. Initally we calculate the number of Pb\SP{2+} that have 6 fold coordination with the nearest I\SP{-} in a cutoff of 0.40 nm and mark them as [PbI\SB{6}]\SP{4-} octahedra. 

2. Secondly, we identify the [PbI\SB{6}]\SP{4-} octahedra that have a coordination number of six with neighbouring [PbI\SB{6}]\SP{4-} octahedra within a distance cutoff of 0.72 nm. This distance cutoff is taken according to the first coordination sphere of Pb\SP{2+} ions in the cubic perovksite phase of MAPI as shown in Fig. \ref{fig:gor}. 

3. Then we count the number of those particular octahedra (displayed in red color in Fig. \ref{fig:Perovskite-corner-sharing}a) that are sharing six I\SP{-} corners with the neighbours and the angle between Pb-I-Pb (as shown in Fig.\ref{fig:Perovskite-corner-sharing}a) of these octahedra is more that 150\degree. The value of this angle is selected from the histrogram of Pb-I-Pb angles displayed in Fig. \ref{fig:Perovskite-corner-sharing}c. This histogram is calculated from a 10ns trajectory of a supercell made of 7x7x7 unit cells of cubic MAPI at 410K and atmospheric pressure in the isothermal-isobaric ensemble as described in the Methods section. These octahedra form the core of the perovksite crystal.

4. The total number of perovskite corner-sharing octahedra are calculated as sum of core (shown in red color in Fig. \ref{fig:Perovskite-corner-sharing}) and their neighbours (shown in green colors in Fig. \ref{fig:Perovskite-corner-sharing}). 

\newpage

\section*{Simulation-(B)}
%\begin{figure} [H] 
%  \includegraphics[width=160mm,scale=1.0]{../../Publish_scripts/Nucleation_pics_all/Initial_cluster_fraction_180.png}
%  \caption{Size of the largest cluster as a function of simulation time.}
%  \label{fig:initial_cluster}
%\end{figure}
%\begin{figure} [H]
%  \includegraphics[width=160mm,scale=1.0]{../../Publish_scripts/Nucleation_pics_all/transition_180.pdf}
%  \caption{Edge-sharing and perovskite corner-sharing octahedra as function of simulation time. Thick lines are the running average of the data and are shown in order to guide the eyes.}
%  \label{fig:transition}
%\end{figure}

\begin{figure}[H]
  \includegraphics[width=\linewidth]{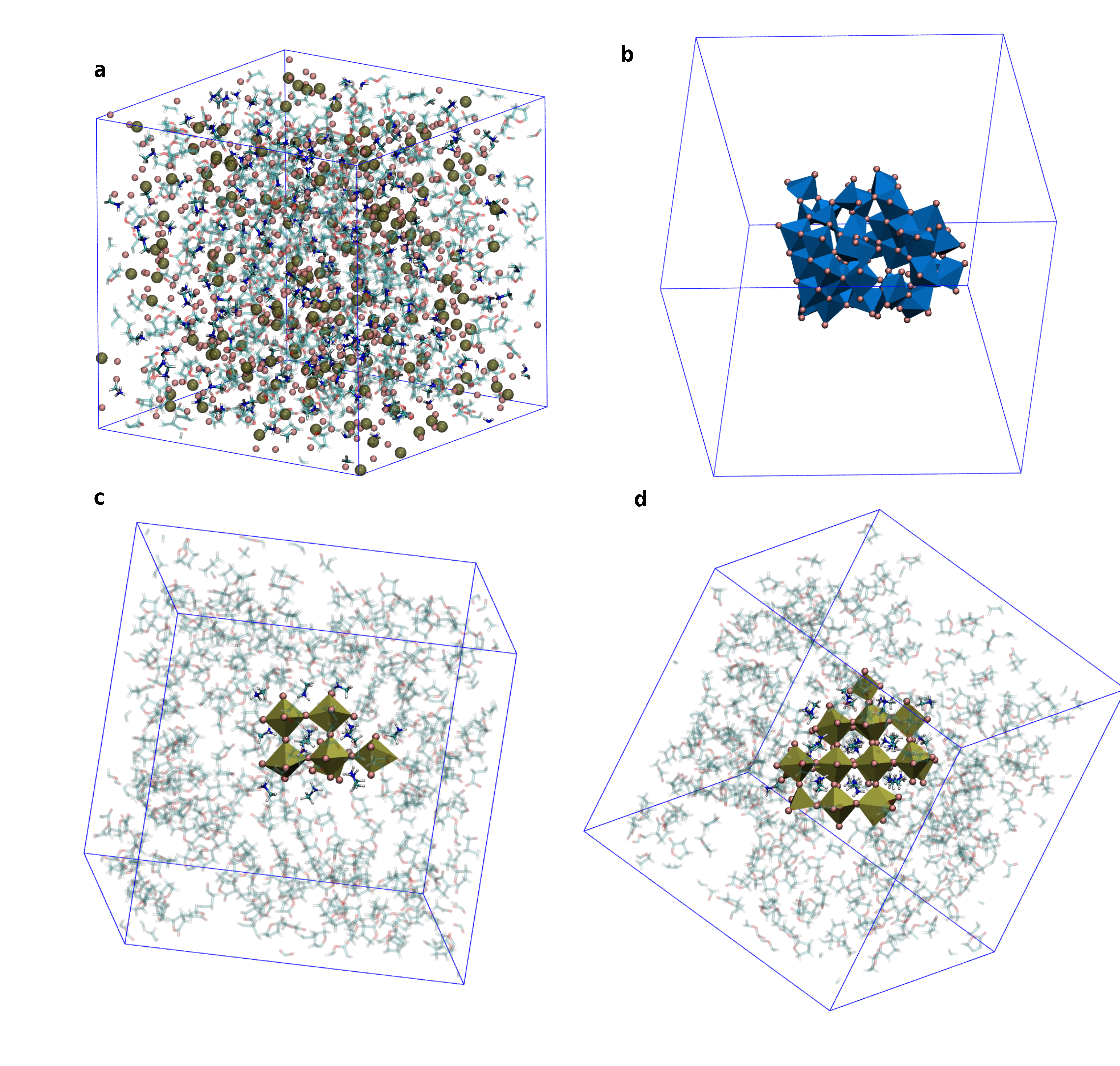}
  \caption{Illustration of the full nucleation pathway for simulation-(B). Pb-I complexes are shown as golden and blue polyhedra with Pb\SP{2+} in the center and I\SP{-} on the corners. I\SP{-} is shown as pink spheres. MA\SP{+} ions are shown with ball and sticks. Snapshot (a) represents the initial solution of MA\SP{+}, I\SP{-} and Pb\SP{2+} in GBL. GBL molecules are respresented with ball and stick model and shown as semi-transparent to visualize the random distribution of Pb\SP{2+}, I\SP{-} and MA\SP{+} in solution. Snapshot (b) represents the intermediate cluster formation of edge-sharing [PbI\SB{6}]\SP{4-} octahedra. Snapshot (c) shows the first perovskite nucleus observed in the solution. Snapshot (d) shows the largest perovskite crystal in this simulations. All the images of these snapshots are generated with VMD-1.9.2.\cite{Humphrey1996}}
  \label{fig:full}
\end{figure}

\newpage
\bibliographystyle{naturemag}
\bibliography{1st_paper.bib}